\documentclass[a4paper,fleqn,usenatbib]{mnras}

\usepackage[T1]{fontenc}
\usepackage{ae,aecompl}

\usepackage{graphicx}	
\usepackage{amsmath}	
\usepackage{amssymb}	
\usepackage{subfigure}
\usepackage{multirow}
\usepackage{longtable}

\newcommand{\dinamo}{{\sc dinamo}}
\newcommand{\pcc}{\,{\rm cm}^{-3}}

\newcommand{\um}{\, {\rm \mu m}}

\newcommand{\kel}{\, {\rm K}}

\newcommand{\msun}{\, {\rm M}_\odot}

\newcommand{\pc}{\, {\rm pc}}

\newcommand{\yr}{\, {\rm yr}}
\newcommand{\kms}{\, {\rm km \, s^{-1}}}

\title[Dust formation in supernovae]{Constraining early-time dust formation in core-collapse supernovae}

\author[Priestley et al.]{
F. D. Priestley$^{1}$,
A. Bevan$^{2}$,  
M. J. Barlow$^{2}$ 
and I. De Looze$^{2,3}$
\\
$^{1}$School of Physics and Astronomy, Cardiff University, Queen's Buildings, The Parade, Cardiff CF24 3AA, UK \\
$^{2}$Department of Physics and Astronomy, University College London, Gower Street, London WC1E 6BT, UK\\
$^{3}$Sterrenkundig Observatorium, Ghent University, Krijgslaan 281 - S9, 9000 Gent, Belgium\\
}

\date{Accepted XXX. Received YYY; in original form ZZZ}

\pubyear{2020}

\begin{document}
\label{firstpage}
\pagerange{\pageref{firstpage}--\pageref{lastpage}}
\maketitle

\begin{abstract}

  There is currently a severe discrepancy between theoretical models of dust formation in core-collapse supernovae (CCSNe), which predict $\gtrsim 0.01 \msun$ of ejecta dust forming within $\sim 1000$ days, and observations at these epochs, which infer much lower masses. { We demonstrate that, in the optically thin case, these low dust masses are robust despite significant observational and model uncertainties. For a sample of 11 well-observed CCSNe, no plausible model reaches carbon dust masses above $10^{-4} \msun$, or silicate masses above $\sim 10^{-3} \msun$. Optically thick models can accommodate larger dust masses, but the dust must be clumped and have a low ($<0.1$) covering fraction to avoid conflict with data at optical wavelengths. These values are insufficient to reproduce the observed infrared fluxes, and the required covering fraction varies not only between SNe but between epochs for the same object. The difficulty in reconciling large dust masses with early-time observations of CCSNe, combined with well-established detections of comparably large dust masses in supernova remnants, suggests that a mechanism for late-time dust formation is necessary.}

\end{abstract}

\begin{keywords}
dust, extinction -- supernovae: general
\end{keywords}



\section{Introduction}

Dust production by core-collapse supernovae (CCSNe) is commonly invoked to explain the large dust masses seen in some high-redshift galaxies \citep{dunne2003,gall2011,gall2018}, requiring a dust yield per SNe of $0.1-1.0 \msun$ according to galaxy evolution models \citep{morgan2003,dwek2007}. Far-infrared (IR) observations of nearby supernova remnants (SNRs) (e.g. \citealt{matsuura2015,delooze2017,temim2017,chawner2019,delooze2019}) have found that many objects studied have dust masses close to or within this range. However, due to confusion with emission from interstellar dust, these far-IR studies of SNRs are generally limited to objects where the ejecta dust has been heated above the interstellar grain temperature, or where the SNR emission can be morphologically distinguished from the foreground/background. Additionally, in old ($\sim 100-1000 \yr$) SNRs the dust has cooled to low enough temperatures that it emits weakly even in the far-IR, meaning even large dust masses may be undetectable with current instruments. In a survey of 71 Galactic SNRs, \citet{chawner2019} could confirm ejecta dust in only four objects, all of which were pulsar wind nebulae, so assuming the dust masses for those objects are representative of the entire sample is not necessarily warranted. { An extended study \citep{chawner2020} found 39 SNRs with associated dust emission, but only one for which this emission could be confidently associated with ejecta dust.} Additionally, while the majority of CCSNe are of type IIP \citep{smith2011}, of SNRs with confirmed ejecta dust emission only the Crab Nebula has been reliably identified as this type \citep{smith2013}. As such, while some CCSNe can certainly produce $\gtrsim 0.1 \msun$ of dust, whether all CCSNe do is currently an open question.

Theoretical models of dust formation in SNe have generally predicted dust masses broadly in agreement with the observed range of $\sim 0.01 - 1.0 \msun$ (e.g. \citealt{nozawa2003,nozawa2010,sarangi2015,bocchio2016,sluder2018}), suggesting that extrapolation from the small number of well-studied SNRs to all CCSNe may be justified. However, when applied to individual objects, these models often find substantial disagreement - \citet{nozawa2010} predicted a dust mass of $0.17 \msun$ for Cas A (and only $0.08 \msun$ at the present day), compared to recent estimated values of $\sim 0.6 \msun$ \citep{delooze2017,bevan2017,priestley2019}, while the predicted mass for the Crab Nebula of $0.25 \msun$ of \citet{bocchio2016} is substantially larger than the $0.03-0.05 \msun$ found by \citet{delooze2019}. The majority of SN dust formation models also predict typical grain sizes of $0.01 \um$ or lower, compared to substantial observational evidence for the presence of micron-sized grains (e.g. \citealt{gall2014,wesson2015,bevan2016,priestley2020}). While the many unknowns entering the calculations, such as progenitor mass, metallicity and the degree of ejecta clumping and mixing, mean that these issues may well be resolvable, it does cast some doubt on the reliability of model predictions.

One prediction common to virtually all dust formation models is that the vast majority of the dust mass should form within $\sim 1000$ days \citep{sarangi2015,sluder2018}, as beyond this point the falling density and temperature of the ejecta should make further dust growth difficult, if not impossible. This is in contrast to the inferred growth of the dust mass in SN 1987A, where observations of both the optical/IR emission \citep{wesson2015} and line profile asymmetries \citep{bevan2016,bevan2018} at multiple epochs suggest that the majority of the dust mass formed at least ten years after explosion. \citet{dwek2015} argued that the infrared emission from SN 1987A is consistent with a much higher dust mass at early times if the dust is optically thick at these wavelengths (see also \citealt{dwek2019}). Compilations of dust masses for multiple objects \citep{gall2014,szalai2019} have found no evidence for non-interacting CCSNe forming $\gtrsim 0.1 \msun$ of dust before day $1000$, which would require the optically thick argument to apply to nearly every case if dust formation models are correct. Note that for several objects the dust mass estimates accounted for optical depth effects (e.g. \citealt{kotak2009,fabbri2011}), as did \citet{wesson2015} themselves.

In deriving dust masses from optical and infrared photometric data, the studies mentioned above generally attempt to fit the available data with models ranging from simple blackbody fits to the mid-IR fluxes, to self-consistent radiative transport models of dust heating in the ejecta, including the optical and ultraviolet (UV) emission. All approaches are subject to potentially significant systematic errors and uncertainties; in particular, determining the `dust' spectral energy distribution (SED) to fit requires dealing with both the underlying SNe flux, which is not necessarily well-described by a blackbody \citep{pejcha2015}, and potential contamination by line emission, such as the CO vibrational band at $4.6 \um$ which is often detected in CCSNe (e.g. \citealt{kotak2005}). { Mid-IR observations are also not sensitive to emission from cold ($\lesssim 50 \kel$) dust grains, which make up the majority of the mass in most observed SNRs. In this paper, we show that despite the above issues, large dust masses cannot realistically be present at early times in SNe unless the dust is optically thick in the IR. We further demonstrate that consistently reproducing the optical and IR behaviour of SNe with optically thick dust appears to require contradictory assumptions about the geometry and/or significant fine-tuning. This casts doubt as to whether theoretical predictions can be reconciled with observation, and suggests dust formation models may need to be reevaluated.}

\section{Optically thin dust}
\label{sec:optthin}

\subsection{Method}

{ Dust masses in SNe are commonly determined by fitting a modified blackbody function, with a minimum of two parameters (mass and temperature), to the observed mid-IR SED. This assumes that the IR fluxes are entirely due to dust emission, which is often clearly not the case, and that the best-fit temperature is a good representation of the mass-weighted average dust temperature. Even without contamination by SN/molecular emission, the much higher emissivity per unit mass of hotter dust grains means that the best-fit blackbody temperature will be biased towards the highest values, leading to potentially large underestimates of the total mass present (e.g. \citealt{shetty2009,priestley2020b}). We instead estimate dust masses using a different method which avoids both of these issues, to investigate whether the large values ($\gtrsim 10^{-2} \msun$) required by models can reasonably be present at early times.}

We first calculate IR SEDs adopting a simple physical model of the SN ejecta. We assume the dust is heated by the radiation field from the SN, which is represented as a point source, with all the dust located at a single distance. We determine the SN luminosity and spectrum by fitting a blackbody to optical/UV data. While SN spectra show deviations from blackbodies, as noted above, for the purposes of dust heating this assumption is acceptable as long as the total luminosity and spectral shape are reasonably similar. For some SNe, extrapolating the blackbody fit into the mid-IR leads to fluxes larger than those observed by Spitzer, so we are potentially overestimating the heating rate. However, given that the heating is dominated by the optical/UV flux, which is well represented by a blackbody, the effect is small.

We assume the dust expands at a constant velocity from SN outburst to obtain the radius, which then determines the incident radiation field. Ejecta velocities in CCSNe are generally in the range $\sim 10^3-10^4 \kms$, with regions forming dust expected to be towards the lower end of this range (e.g. \citealt{nozawa2010}). We use a value of $3000 \kms$ for all objects, which is larger than values found for the dust-forming ejecta in IIP SNe (which make up most of our sample) and SN1987A \citep{matsuura2017}. Higher velocities lead to more dilute radiation fields, lower dust temperatures and thus higher dust masses. For comparison, the IIP model presented in \citet{nozawa2010} has a maximum velocity in the dust-forming region of $1500 \kms$, down to a few hundred $\kms$ in the inner core. Stripped-envelope objects such as IIb SNe may have even higher velocities, up to $\sim 10^4 \kms$ - for these cases we investigate the effect of a higher expansion velocity, but for typical objects our value is likely to be comfortably larger than the highest expected.

Having determined the incident radiation field on the dust, we then calculate the dust SED for a given grain composition and size using \dinamo{} \citep{priestley2019}, which calculates the temperature distribution for each grain size, accounting for stochastic heating effects, and returns the dust emission. We investigate both silicate and carbon grains, using the optical constants and physical properties listed in Table \ref{tab:dustprop} - we use Mg$_2$SiO$_4$ and amorphous carbon grains, predicted by most dust formation models to be the most abundant species \citep{sarangi2015,bocchio2016,sluder2018}. The optical constants are combined to extend the covered wavelength ranges as described in \citet{priestley2019}. Theoretical dust formation studies have frequently found approximately log-normal grain size distributions \citep{nozawa2003,sarangi2015,biscaro2016}, but both the peak and the extent of the distributions are heavily dependent on model assumptions. Again in the interests of maximising our dust { masses}, we use a single grain size of $1.0 \um$, in line with several observational studies \citep{gall2014,wesson2015,bevan2016,priestley2020} finding evidence for grains of this size in SNRs. { Larger grains are generally cooler and therefore less efficient emitters, requiring larger dust masses to fit a given IR flux.}

{ Once we have calculated the dust SED, the model flux depends only on the total dust mass (in the optically thin case). Given the potential contamination of the observed fluxes by various non-dust sources, we do not attempt to fit this as a free parameter. Instead, we determine the maximum value for which all model fluxes are below the observed values, taking into account the measurement uncertainties. While this method does not necessarily produce a good fit, it does return a maximum dust mass given our model assumptions, which is insensitive to the underlying IR SED of the SN itself or any potential line emission. As we have calculated, rather than fit, the dust temperature, it is also unaffected by temperature bias - larger masses of cold dust can only be present if they are strongly shielded from the SN radiation, in which case our assumption of optically thin emission is likely to be inaccurate anyway.}

\begin{table}
  \centering
  \caption{Dust species and their adopted densities $\rho_g$, sublimation temperatures $T_{\rm sub}$ and references for the optical constants. References: (1) \citet{jaeger2003} (2) \citet{laor1993} (3) \citet{zubko1996} (4) \citet{uspenskii2006}.}
  \begin{tabular}{cccc}
    \hline
    Dust species & $\rho_g$/${\rm g} \pcc$ & $T_{\rm sub}$/$\kel$ & $n$-$k$ \\
    \hline
    Mg$_2$SiO$_4$ & $2.5$ & $1500$ & (1),(2) \\
    ACAR & $1.6$ & $2500$ & (3),(4) \\
    \hline
  \end{tabular}
  \label{tab:dustprop}
\end{table}

\begin{figure}
  \centering
  \includegraphics[width=\columnwidth]{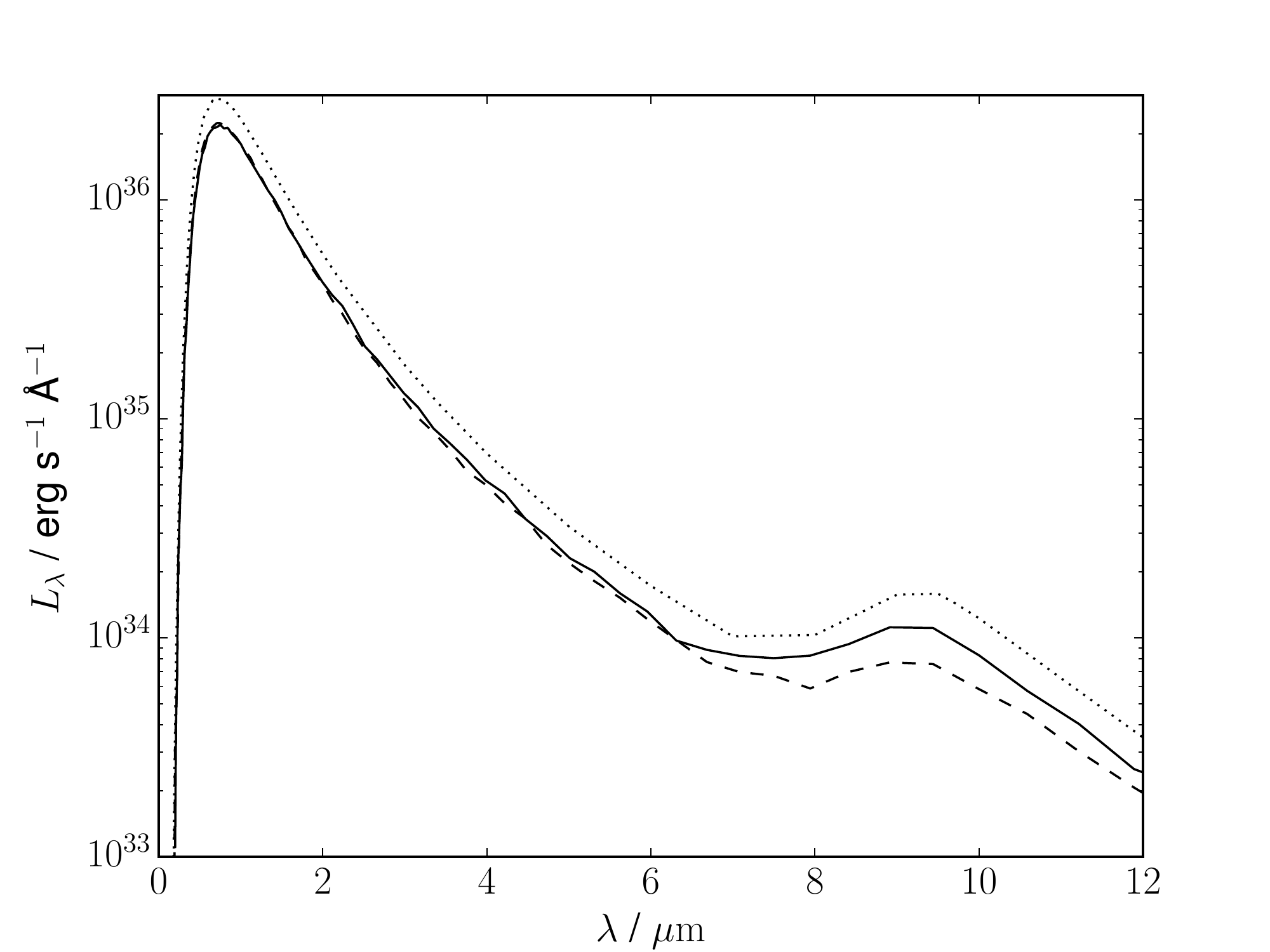}
  \caption{{\sc mocassin} model SEDs for diffuse source (solid line) and shell-central source (dashed line) geometries, and the dust SED calculated by \dinamo{} for the equivalent dust shell parameters combined with the blackbody central heating source (dotted line). The input blackbody radiation is reprocessed into the IR by the dust.}
  \label{fig:modeltest}
\end{figure}

Our model setup is not obviously applicable to the ejecta of a CCSNe, where the dust and the radioactive elements fueling the emission are presumably both distributed throughout the ejecta, with a resulting distribution of radiation field intensities. However, we find that the resulting dust emission is not hugely sensitive to the chosen geometry. Figure \ref{fig:modeltest} shows the output SEDs for a thin shell surrounding a point source, and a constant density sphere with a diffuse radiation field, calculated using the Monte Carlo radiative transfer code {\sc mocassin} \citep{ercolano2003,ercolano2005,ercolano2008}. The radius (of either the sphere or the shell) is $10^{16}$ cm, the silicate dust mass is $10^{-5} \msun$ and the source is assumed to be a blackbody with luminosity $10^{41} \, {\rm erg \, s^{-1}}$ and temperature $4000 \kel$ (parameters comparable to a CCSNe $\sim 100$ days post-explosion). Our shell model has a thickness of $10^{15}$ cm. We also show the dust SED calculated using \dinamo{} for the same parameters, combined with the blackbody heating source. The \dinamo{} model flux is a factor of $\sim 2$ higher than the {\sc mocassin} models in the mid-IR region. This is due to \dinamo{} not accounting for attenuation of the radiation field by the dust. Fitting a blackbody to the {\sc mocassin} diffuse SED { in the optical/UV region} returns a reduced luminosity of $7.3 \times 10^{40} \, {\rm erg \, s^{-1}}$ and temperature $3900 \kel$ - using these parameters and the {\sc mocassin} mid-IR flux, { our method returns a dust mass of $1.5 \times 10^{-5} \msun$, consistent with (and somewhat larger than) the input value.}

In order to determine dust { masses using this method}, we require near- to mid-IR data to constrain the dust emission, and optical/UV fluxes from the same epoch to calculate the heating rate and dust temperatures. We searched the Open Supernova Catalog (OSC) \citep{guillochon2017} for non-interacting CCSNe (i.e. types Ib, Ic, IIb, IIL and IIP) with Spitzer IRAC observations beyond $\sim 100$ days and contemperaneous optical observations in at least three bands, or a clear exponential decay so that we can extrapolate to later epochs. We also utilise Spitzer data not included in the OSC from various other sources. We exclude interacting supernovae from our sample, as in these cases the majority of the IR emission presumably comes from shock-heated dust in the circumstellar material (CSM). This gives a sample of 11 CCSNe, with 35 separate epochs of mid-IR data with adequate optical/UV observations spanning from 58 to 1351 days post-explosion. We determine fluxes in the optical/UV bands by interpolating between the two nearest observations in time, or extrapolating linearly in magnitude if necessary (i.e. an exponential decay). We deredden using the reddening from \citet{schlafly2011} and the extinction law of \citet{fitzpatrick1999}. Finally, we use measured distances to the SN host galaxies (listed in Table \ref{tab:snprop}) in order to convert to the luminosity SED, allowing us to apply the method described above to determine { dust masses} for each epoch. Basic properties for each SN in our sample are listed in Table \ref{tab:snprop}, and the resulting dust masses for each epoch are given in Table \ref{tab:photometry}. Our derived SEDs and uncertainties at each epoch are given in Tables \ref{tab:sed} and \ref{tab:sederr}.

\begin{table*}
  \centering
  \caption{Name, type, distance and reddening value for each SN in our sample. Reddening values are from \citet{schlafly2011}.}
  \begin{tabular}{ccccc}
    \hline
    Name & Type & $d / {\rm Mpc}$ & E(B-V) & Ref. (type and distance) \\
    \hline
    SN2004et & IIP & $7.7$ & $0.29$ & \citet{fabbri2011,anand2018} \\
    SN2013am & IIP & $18.5$ & $0.02$ & \citet{yaron2012,nakano2013} \\
    SN2004A & IIP & $20.3$ & $0.01$ & \citet{hendry2006} \\
    SN2013aw & IIP & $8.8$ & $0.02$ & \citet{henden2012,szalai2019} \\
    SN2004dj & IIP & $3.5$ & $0.03$ & \citet{vinko2006} \\
    SN2003gd & IIP & $9.3$ & $0.06$ & \citet{hendry2005} \\
    SN2009E & II Pec & $37.5$ & $0.02$ & \citet{pastorello2012,szalai2019} \\
    SN2007it & IIP & $11.7$ & $0.10$ & \citet{andrews2011} \\
    SN2013ej & IIP/L & $9.5$ & $0.06$ & \citet{mauerhan2017} \\
    SN2013df & IIb & $16.6$ & $0.02$ & \citet{szalai2016} \\
    SN2011dh & IIb & $7.7$ & $0.03$ & \citet{arcavi2011,helou2013} \\
    \hline
  \end{tabular}
  \label{tab:snprop}
\end{table*}

\subsection{Results}

\begin{table*}
  \centering
  \caption{Adopted time post-explosion and dust { masses} for carbon and silicate grains, for each epoch of IR photometry. A list of the obtained SEDs and uncertainties can be found in Tables \ref{tab:sed} and \ref{tab:sederr}.}
  \begin{tabular}{ccccc}
    \hline
    Name & $t / {\rm d}$ & $M_{\rm car} / 10^{-6} \msun$ & $M_{\rm sil} / 10^{-3} \msun$ & Ref. \\
    \hline
    \multirow{5}{*}{SN2004et} & 58 & $0.0$ & $0.0$ & \citet{li2005,maguire2010} \\
    & 294 & $1.7$ & $0.3$ & \citet{fabbri2011,faran2014} \\
    & 354 & $2.3$ & $0.7$ & \\
    & 398 & $4.1$ & $1.7$ & \\
    & 458 & $9.3$ & $1.7$ & \\
    \hline
    \multirow{3}{*}{SN2013am} & 351 & $1.1$ & $1.6$ & \citet{zhang2014,rubin2016} \\
    & 377 & $1.7$ & $5.2$ & \citet{tinyanont2016} \\
    & 501 & $3.2$ & $273.4$ & \\
    \hline
    \multirow{3}{*}{SN2004A} & 238 & $2.4$ & $0.3$ & \citet{nakano2004,hendry2006} \\
    & 436 & $16.4$ & $4.1$ & \citet{szalai2013} \\
    & 554 & $56.0$ & $3.8$ & \\
    \hline
    \multirow{2}{*}{SN2012aw} & 357 & $1.5$ & $11.3$ & \citet{bose2013,dallora2014} \\
    & 486 & $4.8$ & $286.6$ & \citet{rubin2016,szalai2019} \\
    \hline
    \multirow{8}{*}{SN2004dj} & 92 & $0.0$ & $0.0$ & \citet{zhang2006,szalai2011} \\
    & 236 & $1.2$ & $0.07$ & \citet{meikle2011,guillochon2017} \\
    & 445 & $15.4$ & $0.5$ & \\
    & 600 & $38.2$ & $1.0$ & \\
    & 822 & $68.4$ & $1.4$ & \\
    & 975 & $69.1$ & $1.4$ & \\
    & 1210 & $63.0$ & $1.8$ & \\
    & 1351 & $74.2$ & $2.0$ & \\
    \hline
    SN2003gd & 410 & $18.8$ & $11.4$ & \citet{hendry2005,meikle2007} \\
    \hline
    SN2009E & 547 & $68.7$ & $926.2$ & \citet{lennarz2012,szalai2019} \\
    \hline
    \multirow{4}{*}{SN2007it} & 350 & $3.6$ & $0.4$ & \citet{andrews2011,anderson2014} \\
    & 560 & $78.5$ & $13.2$ & \\
    & 718 & $777.1$ & $10466.3$ & \\
    & 942 & $1414.9$ & $2620.9$ & \\
    \hline
    \multirow{3}{*}{SN2013ej} & 237 & $4.9$ & $6.7$ & \citet{huang2015,yuan2016} \\
    & 260 & $6.0$ & $13.5$ & \citet{mauerhan2017} \\
    & 439 & $24.8$ & $420.5$ & \\
    \hline
    \multirow{2}{*}{SN2013df} & 264 & $2.8$ & $12.2$ & \citet{vandyk2014,moralesgaroffolo2014} \\
    & 291 & $2.5$ & $25.5$ & \citet{maeda2015,tinyanont2016} \\
    \hline
    \multirow{3}{*}{SN2011dh} & 249 & $5.0$ & $4.2$ & \citet{sahu2013,tinyanont2016} \\
    & 279 & $8.3$ & $10.5$ & \\
    & 312 & $6.3$ & $7.3$ & \\
    \hline
  \end{tabular}
  \label{tab:photometry}
\end{table*}

\begin{figure}
  \centering
  \includegraphics[width=\columnwidth]{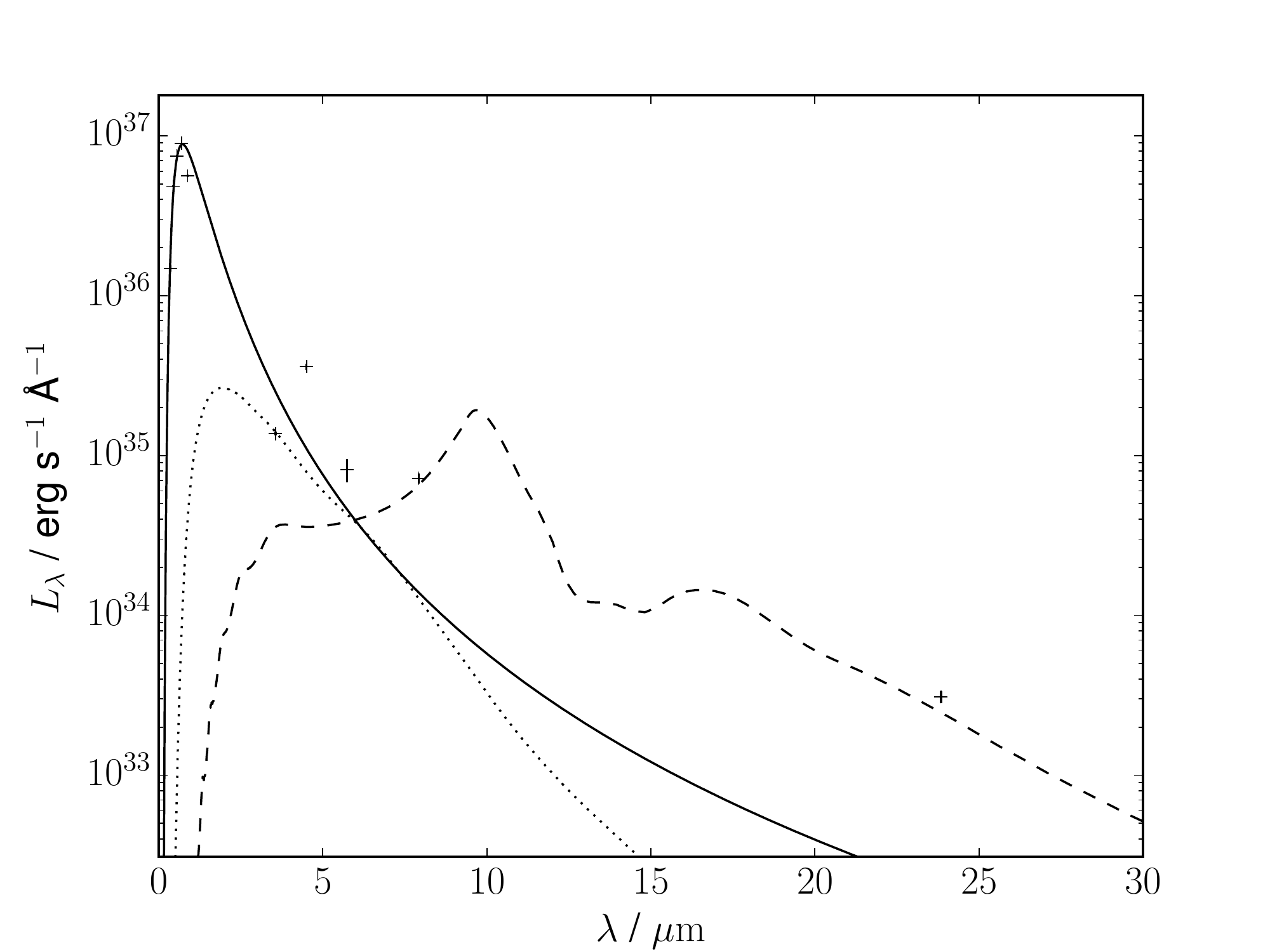}
  \caption{Photometric fluxes for SN2004et at day 294 (black crosses), our blackbody fit to the optical/UV data (solid line) and our upper limit dust SEDs for $1.0 \um$ carbon (dotted line) and silicate grains (dashed line).}
  \label{fig:2004etd294}
\end{figure}

{ Figure \ref{fig:2004etd294} shows the result of our method for SN2004et at day 294. In this case, a silicate dust mass of $3 \times 10^{-4} \msun$ combined with the blackbody fit to the optical data appears to fit the IR observations well, with the exception of the $4.5 \um$ flux, which as previously noted can be contaminated by CO emission. The presence of a silicate feature at $10 \um$ \citep{kotak2009} means that at least some of the newly formed dust must be of this composition. Carbon grains are unable to reproduce the longer wavelength fluxes, as at the grain temperatures produced by our model the resulting flux at shorter wavelengths would be far higher than those observed. The blackbody fit to the optical data already exceeds the flux at $3.6 \um$, highlighting the difficulties in disentangling SN and dust emission at these wavelengths, but as the dust heating is dominated by the much more intense optical radiation the effect on our calculated dust temperatures is minimal. Previous authors \citep{kotak2009,fabbri2011,szalai2019} have found silicate dust masses ranging from $10^{-5}-10^{-4} \msun$ at day 458, the final epoch before signs of interaction with CSM \citep{kotak2009}, compared to our $1.7 \times 10^{-3} \msun$, suggesting our method is returning generously high masses as designed.}

\begin{figure}
  \centering
  \includegraphics[width=\columnwidth]{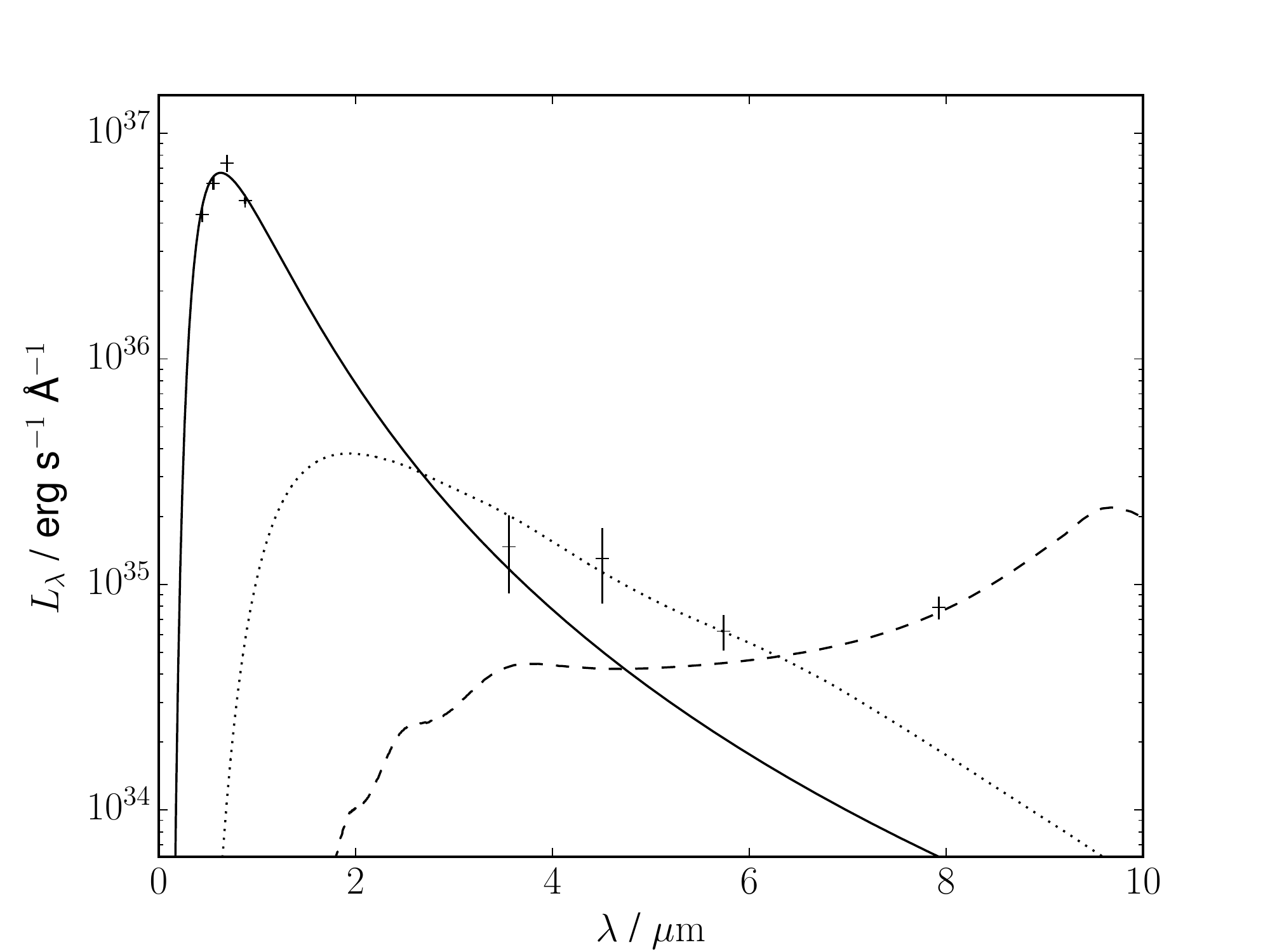}
  \caption{Photometric fluxes for SN2004A at day 238 (black crosses), our blackbody fit to the optical/UV data (solid line) and our upper limit dust SEDs for $1.0 \um$ carbon (dotted line) and silicate grains (dashed line).}
  \label{fig:2004Ad238}
\end{figure}

{ We do not find higher dust masses than previous authors for all SNe. For SN2004A, at day 238 \citet{szalai2013} require $1.6 \times 10^{-3} \msun$ of carbon dust compared to $2.4 \times 10^{-6} \msun$ from our model. The SED, shown in Figure \ref{fig:2004Ad238}, is relatively flat between $6-8 \um$, which \citet{szalai2013} require a blackbody temperature of $\sim 300 \kel$ to fit. The carbon grain temperatures calculated from the SN radiation field are $\gtrsim 1000 \kel$, explaining our far lower dust mass estimate, and are unable to fit the $8 \um$ flux without exceeding the values at shorter wavelengths. Silicate grains, which have lower opacities at optical wavelengths and thus lower temperatures, can reproduce the $8 \um$ flux with $3 \times 10^{-4} \msun$ of dust. \citet{szalai2013} suggest that this emission may originate from pre-existing CSM dust, explaining the unrealistically low temperature for carbon grains. In any case, the mass of newly-formed ejecta dust is clearly below $10^{-3} \msun$.}

\begin{figure}
  \centering
  \includegraphics[width=\columnwidth]{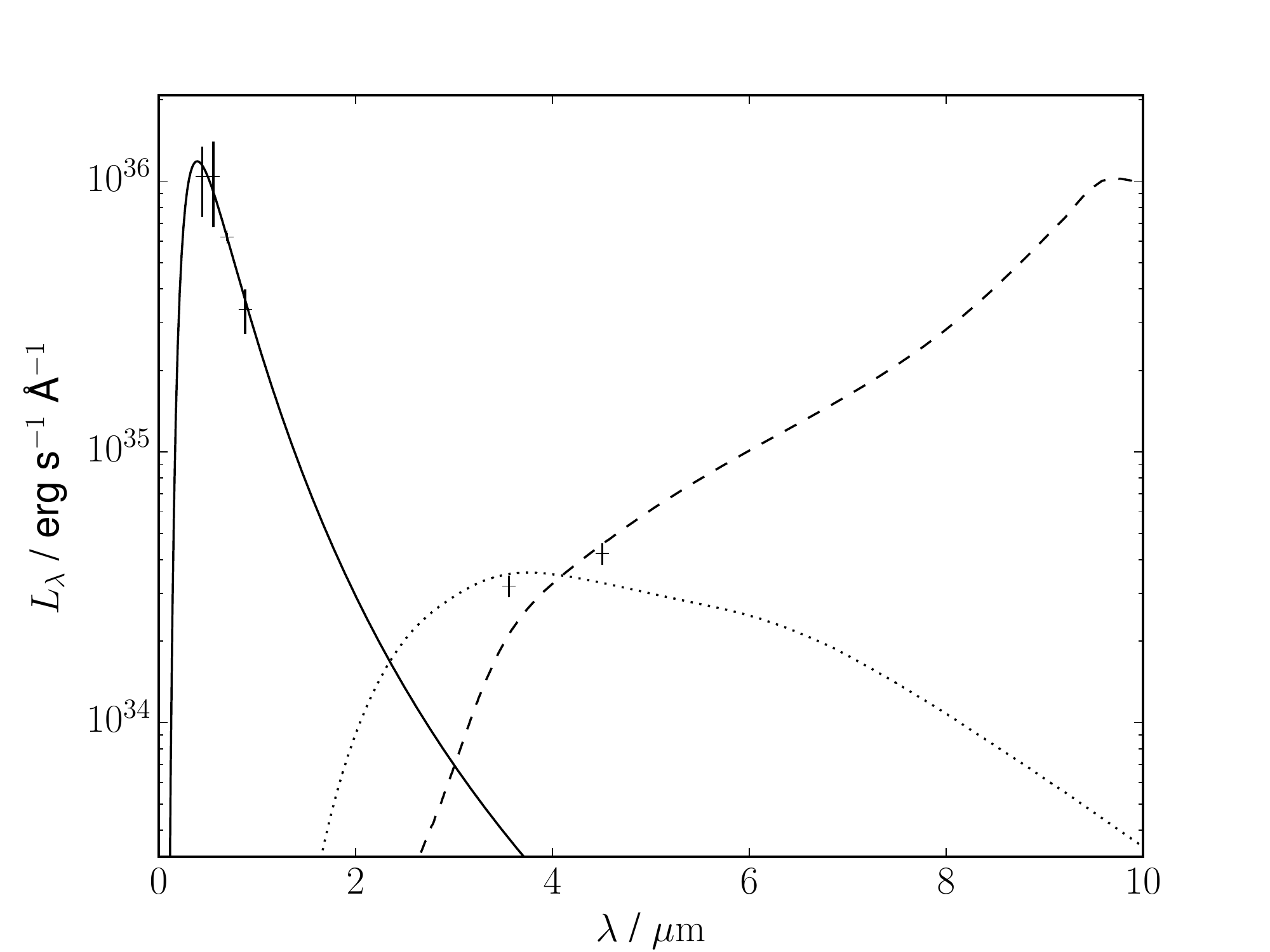}
  \caption{Photometric fluxes for SN2011dh at day 312 (black crosses), our blackbody fit to the optical/UV data (solid line) and our upper limit dust SEDs for carbon (dotted line) and silicate (dashed line) grains.}
  \label{fig:2011dhd312}
\end{figure}

\begin{figure}
  \centering
  \includegraphics[width=\columnwidth]{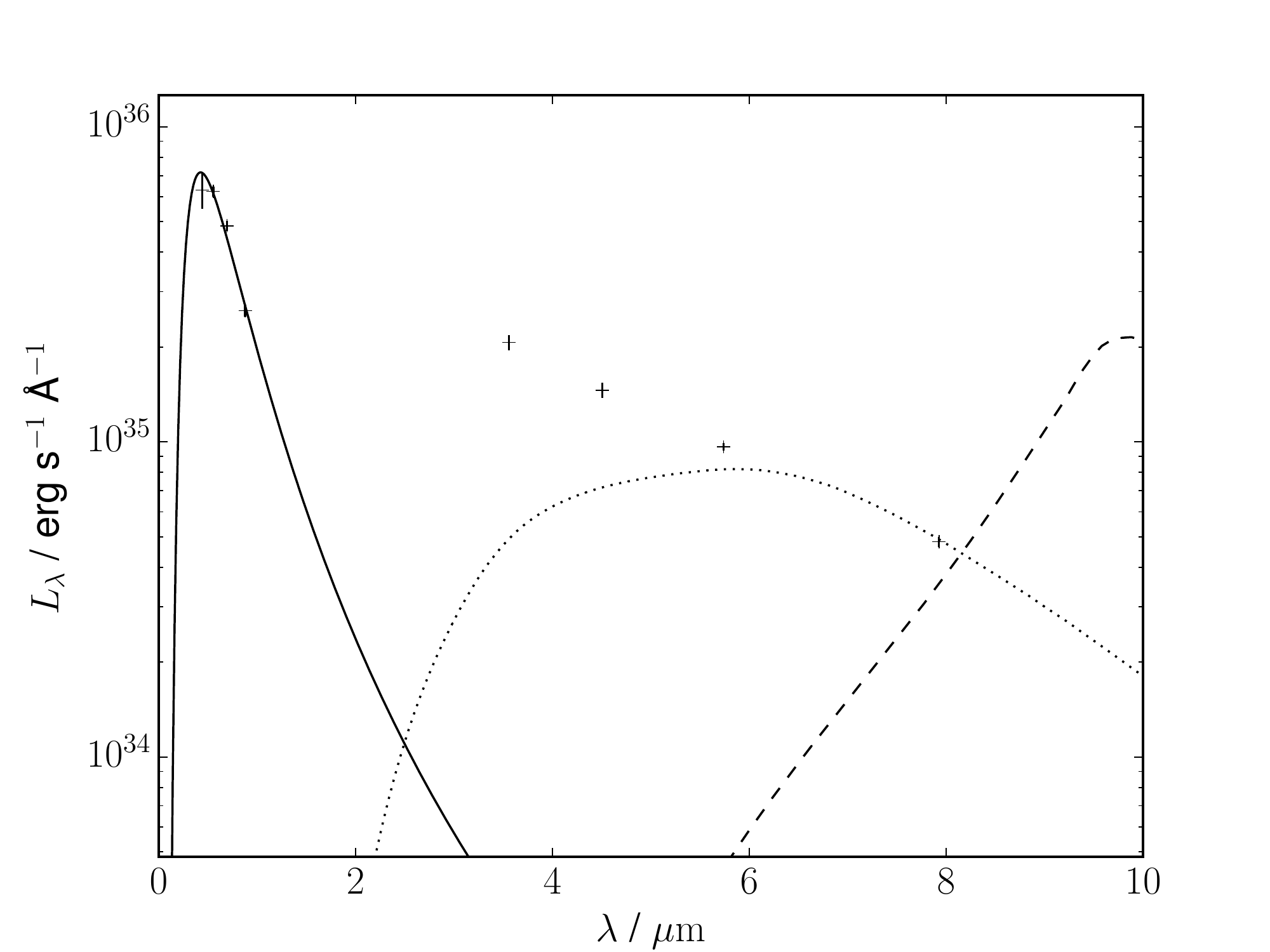}
  \caption{Photometric fluxes for SN2007it at day 560 (black crosses), our blackbody fit to the optical/UV data (solid line) and our upper limit dust SEDs for carbon (dotted line) and silicate (dashed line) grains.}
  \label{fig:2007itd560}
\end{figure}

{ The lower temperatures of silicate grains for a given radiation field can lead to unrealistic values for the dust masses, particularly when IR data beyond $5 \um$ is lacking. Figure \ref{fig:2011dhd312} shows the observed fluxes and our model SEDs for SN2011dh at day 312. While both carbon and silicate grains can reasonably fit the two IRAC data points, the required total IR luminosity for silicate grains is comparable to the SN itself, which we regard as implausible. In this case the silicate dust mass is $7.3 \times 10^{-3} \msun$, larger than the values found by \citet{tinyanont2016} and \citet{szalai2019} but not unreasonable. In some cases the masses required are clearly unphysical, e.g. $10.5 \msun$ for SN2007it at day 718. Smaller grain sizes, or smaller distances from the SN, would reduce these values, but we regard dust masses derived from limited IR data as highly unreliable. Even when observations in all four IRAC bands are available (e.g. SN2007it at day 560, Figure \ref{fig:2007itd560}), the implied flux at $10 \um$ and beyond from silicate grains can be unreasonably large. In general, we find that at the grain temperatures calculated from the optical data, silicate masses above $\sim 10^{-3} \msun$ would require that the dust luminosity is comparable to, or larger than, the SN flux in the optical.}

\begin{figure*}
  \centering
  \subfigure{\includegraphics[width=\columnwidth]{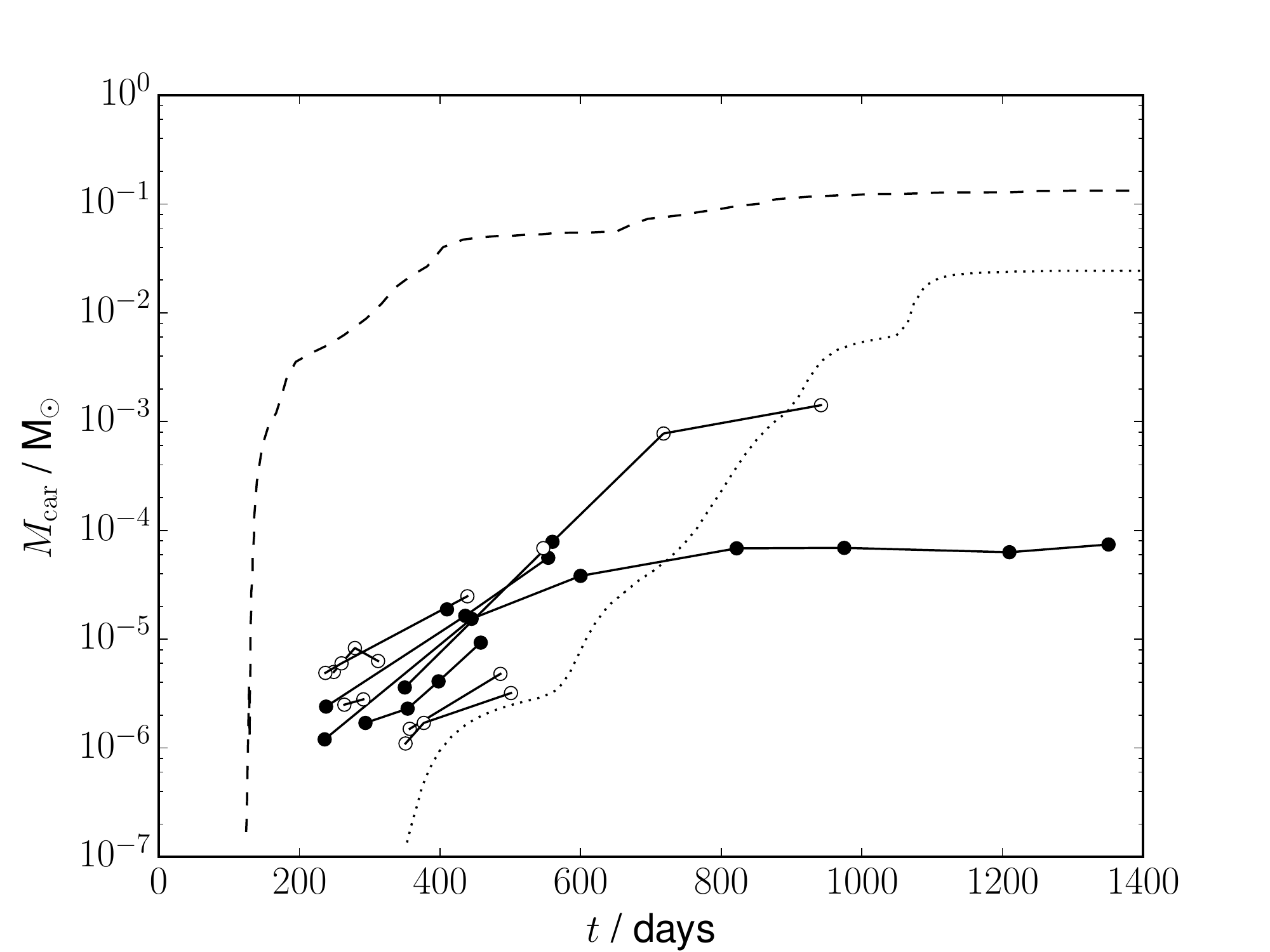}}\quad
  \subfigure{\includegraphics[width=\columnwidth]{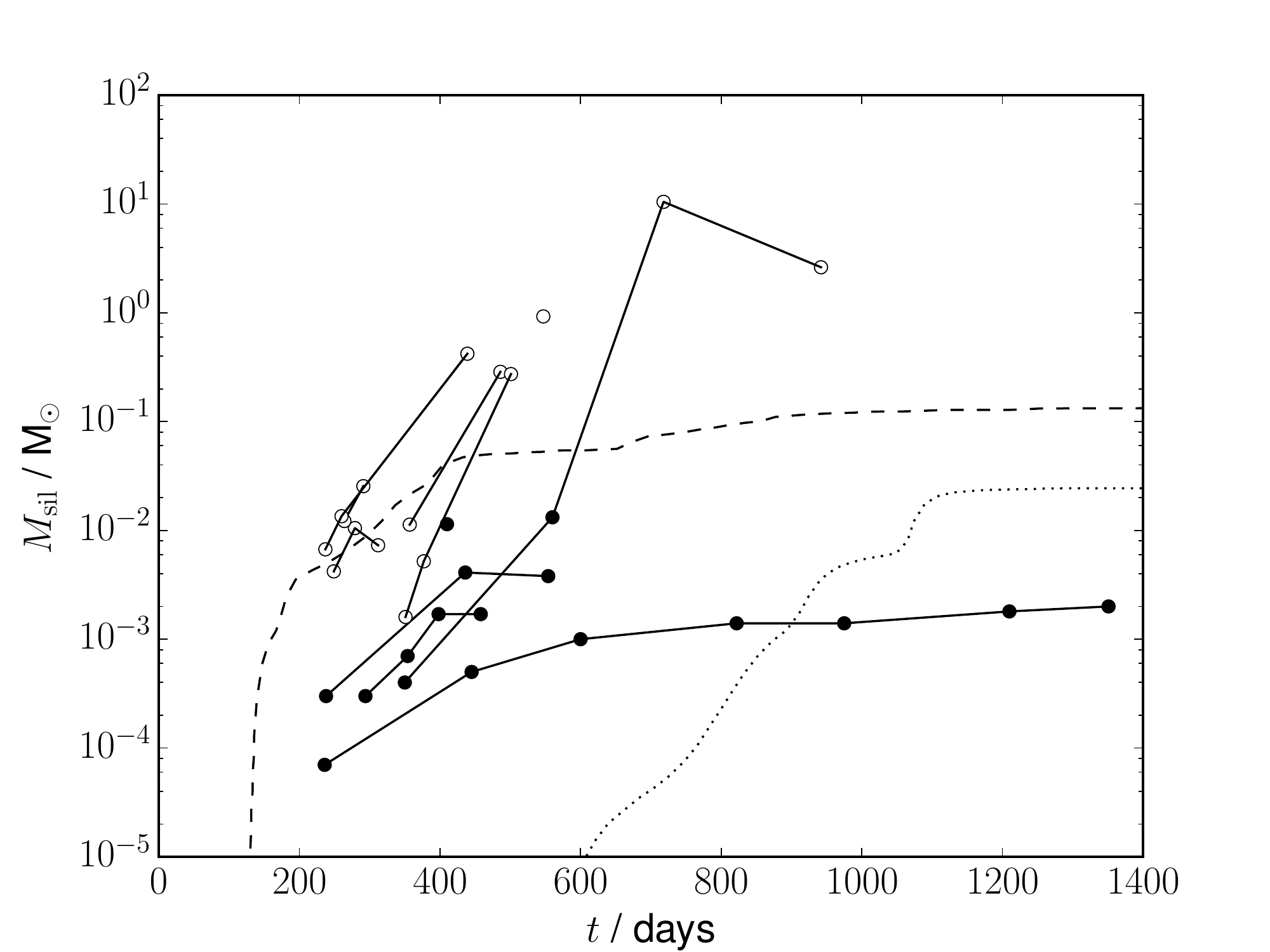}}
  \caption{Carbon (left panel) and silicate (right panel) { dust masses} for each epoch for our CCSN sample, { as listed in Table \ref{tab:photometry}}. Epochs with all four IRAC bands are shown as filled points, those with only $3.6$ and $4.5 \um$ data as unfilled. Epochs from the same supernova are linked with solid lines. Model dust mass predictions from \citet{sarangi2015} are shown for a $19 \msun$ progenitor with homogenous (dotted line) or clumpy (dashed line) ejecta.}
  \label{fig:masstime}
\end{figure*}

Figure \ref{fig:masstime} shows the { dust masses} for our entire sample of CCSNe as a function of time post-explosion, along with two dust formation models for the explosion of a $19 \msun$ star from \citet{sarangi2015}. { These models predict final dust masses an order of magnitude lower than those commonly found in SNRs, so represent a lower limit of sorts.} Excluding epochs with only two IRAC bands available, we find no evidence for the formation of more than $10^{-4} (10^{-2}) \msun$ of carbon (silicate) grains in the first $\sim 1000$ days post-explosion. The two epochs with { data beyond $4.5 \um$ and} a silicate { mass} of $0.01 \msun$ { clearly require implausibly high fluxes at $\ge 10 \um$ for these dust masses to exist, as can be seen in Figure \ref{fig:2007itd560}}, { and the problem is even worse for those epochs with only the two shorter-wavelength IRAC bands (Figure \ref{fig:2011dhd312}). The rapid increase in silicate mass for these epochs between days $200-400$ is due to the decrease in grain temperature drastically reducing the emissivity at short wavelengths, rather than a sign of dust formation, even if the required IR luminosities were realistic.} The maximum realistic silicate mass is of order $10^{-3} \msun$, e.g. for SN2004dj, { where data at $24 \um$ is available}. Even including epochs with only two IRAC bands, the { maximum} mass of carbon grains formed is $\sim 10^{-3} \msun$, in conflict with many (if not all) models of ejecta dust formation. The majority of models presented in \citet{sarangi2015} and \citet{sluder2018} predict $\gtrsim 0.01 \msun$ of dust to have formed by day $600$, which is in conflict with even our silicate { masses}. The late-time ($>600$ days) formation of large masses of carbon dust found by \citet{sarangi2015} in some models { also has no observational support.} Despite the diversity of objects in our sample, the trend between days $200-600$ seems to be relatively uniform, with the { dust mass} increasing by $\sim 2$ orders of magnitude (beyond day 600 only SN2004dj has observations with full IRAC coverage). Theoretical models of dust formation appear to predict at least an order of magnitude more dust at early times than there is observational evidence for.

Our results are broadly consistent with \citet{tinyanont2016} and \citet{szalai2019}, in that most non-interacting CCSNe show no evidence for dust masses above $10^{-3} \msun$ forming within a few years of explosion. While some objects modelled by those authors have inferred dust masses larger than this, as discussed previously this can often be attributed to erroneously low blackbody temperatures obtained by fitting relatively few data points with potentially significant soures of contamination. Our results are also generally in agreement with previous work on the individual objects considered, and when dust masses higher than { ours} have been reported they are still well below the masses predicted by dust formation models. The dust evolution trends reported for SN1987A \citep{wesson2015,bevan2016} and SN2005ip \citep{bevan2019} provide further support for this (SN2005ip is an interacting supernova, but \citet{bevan2019} find that the line asymmetries can be explained entirely by ejecta dust). { If current models of dust formation in SNe are correct, the newly-formed dust must be optically thick in the IR.}

\section{Optically thick dust}
\label{sec:optthick}

{ For dust that is optically thick to its own IR radiation, increasing the mass will not result in a higher IR flux, and models which do not account for radiative transfer effects will only return a lower limit to the true dust mass.} \citet{dwek2015} and \citet{dwek2019} have invoked this argument to claim that the low early-time dust masses measured from IR observations of SN1987A are erroneous, and the IR emission actually comes from a much larger mass ($>0.1 \msun$) of optically thick dust, in better agreement with theoretical predictions \citep{sluder2018}. If the dust is optically thick in the IR, then for any plausible composition it will also be optically thick at optical/UV wavelengths. We consider the effects of this below.

\begin{figure}
  \centering
  \includegraphics[width=\columnwidth]{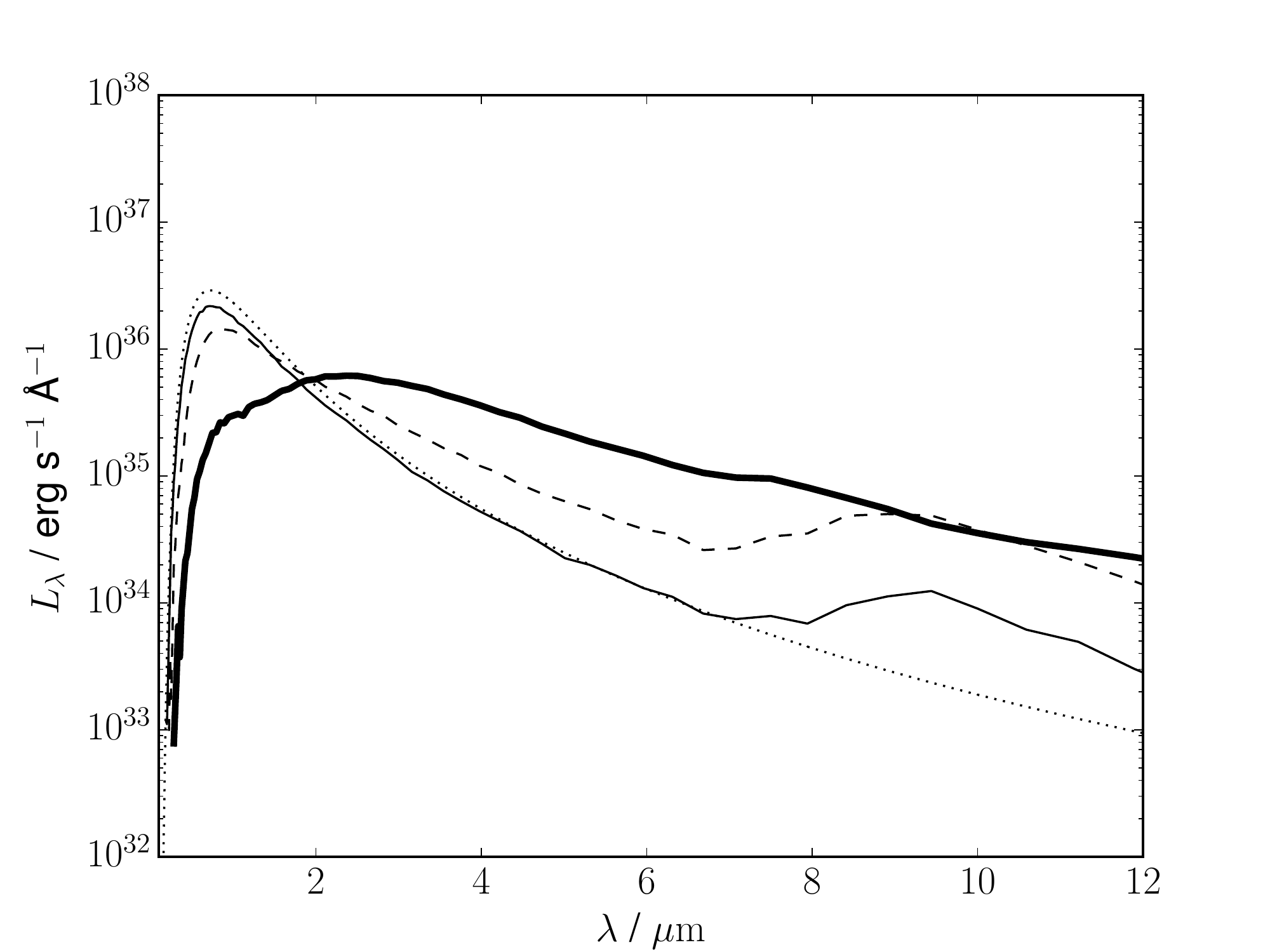}
  \caption{Model SEDs for $10^{-5}$ (thin solid line), $10^{-4}$ (dashed line) and $10^{-3} \msun$ (thick solid line) of uniformly distributed dust, heated by a diffuse blackbody source (dotted line).}
  \label{fig:optthick}
\end{figure}

We generated SEDs using {\sc mocassin} for the uniform sphere model presented in Section \ref{sec:optthin}, shown in Figure \ref{fig:optthick}, with increasing values of the silicate dust mass. Going from $10^{-5}$ to $10^{-4} \msun$, the attenuation in the optical is minimal, and the dust IR flux increases in proportion to the mass. At $10^{-3} \msun$, the silicate dust becomes optically thick beyond $8 \um$ (silicates remain optically thin in the near-IR for longer as the opacity is lower) and the flux no longer increases with dust mass. In addition, the optical emission from the heating source is severely attenuated and reprocessed to IR wavelengths, resembling a blackbody of temperature $\sim 1000 \kel$. For all epochs considered in this paper, the optical observations are well fit by blackbodies with temperatures $\gtrsim 3000 \kel$, which is incompatible with reprocessed emission from optically thick dust. In addition, dust formation models predict that once conditions allow a species to condense out, it does so rapidly ($\sim 10$ days; \citealt{sarangi2015,sluder2018}), which would appear as a sharp drop in the observed flux if the mass formed is large enough to become optically thick. None of the lightcurves for our sample of SNe shows evidence for this - if large masses of optically thick dust are formed in SNe, it must be in clumps, or some alternative geometry where the observer's line of sight is unobscured.

Clumped dust models have frequently been invoked in radiative transfer models of supernova ejecta (e.g. \citealt{sugerman2006,ercolano2007,fabbri2011,wesson2015}), based on both theoretical expectations \citep{arnett1989,wongwathanarat2015} and the necessity of fitting optical and IR fluxes simultaneously. Continuing with the basic model parameters from Section \ref{sec:optthin}, we set the clump radius to be $0.01$ times the radius of the ejecta (i.e. $10^{14} \, {\rm cm}$). We fix the dust mass to $0.01 \msun$ and vary the covering fraction $f_{\rm cov}$ of the ejecta, calculated as $f_{\rm cov} = N_{\rm clump} (r_{\rm clump} / r_{\rm ejecta})^2$ where $N_{\rm clump}$ is the total number of clumps. The clump number density follows from the total dust mass and clump volume. Clumps are distributed randomly within $r_{\rm ejecta}$. { For the clumped models we use a central point source rather than diffuse emission to reduce computational cost, which may produce different results to the more realistic case. For a uniform distribution of clumps as we have assumed, however, the average filling factor seen at each point in the sphere does not vary greatly, so our conclusions are not significantly affected.}

\begin{figure}
  \centering
  \includegraphics[width=\columnwidth]{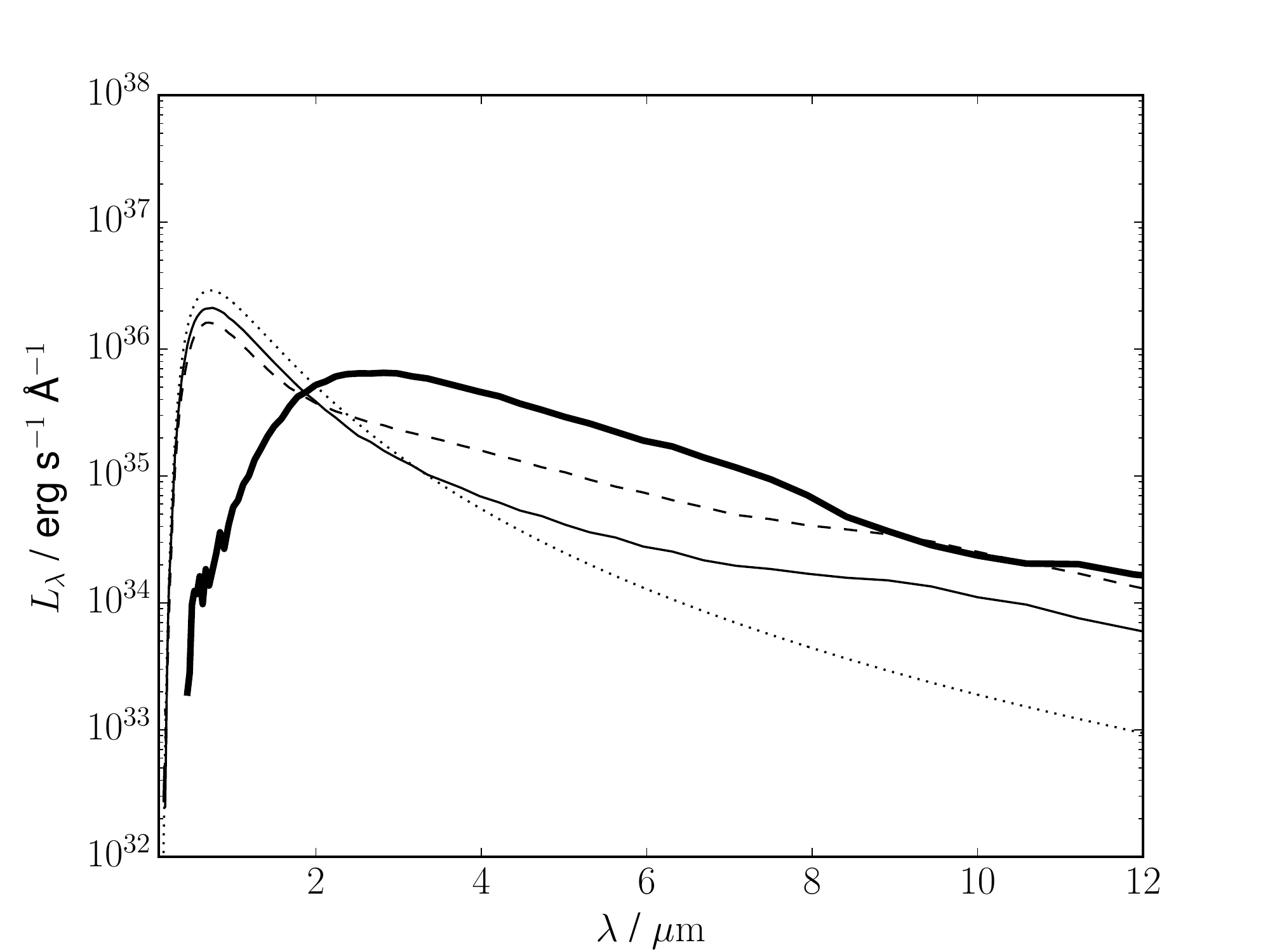}
  \caption{Model SEDs for $10^{-2} \msun$ of dust distributed uniformly (thick solid line), and as clumps with covering fractions of $0.1$ (thin solid line) and $0.5$ (dashed line), and the blackbody heating source (dotted line).}
  \label{fig:clump}
\end{figure}

Figure \ref{fig:clump} shows the SEDs for clumped models with covering fractions $0.1$ and $0.5$, and for a uniform distribution of the same dust mass. Even for the higher covering fraction, the extinction in the optical for clumped models is much lower than for the uniform case, while the near-IR emission is also reduced. We convert the model SEDs into photometric fluxes, applying a $5\%$ uncertainty (typical of the observational data) to each point, and apply our method from Section \ref{sec:optthin} to this mock data, fitting a blackbody to the optical/UV fluxes and using this to determine a dust SED and mass from the mid-IR data. In the smooth case, our dust { mass} is $0.0096 \msun$, very close to the true value - as the assumed heating source is the heavily attenuated blackbody flux, the dust temperature is lower than would be calculated for the input spectrum. For the clumped models, we find dust { masses} of $3 \times 10^{-5}$ and $8 \times 10^{-5} \msun$ for covering fractions of $0.1$ and $0.5$ respectively, orders of magnitude less than the true mass. { It is therefore possible to `hide' large dust masses from detection at early times in optically thick clumps. However, this does {\it not} imply that optically thick clumps are consistent with the observed properties of CCSNe - the inferred silicate masses are lower than those found for the majority of SNe in Section \ref{sec:optthin}, suggesting that the model IR fluxes are below those typically observed.}

\begin{figure}
  \centering
  \includegraphics[width=\columnwidth]{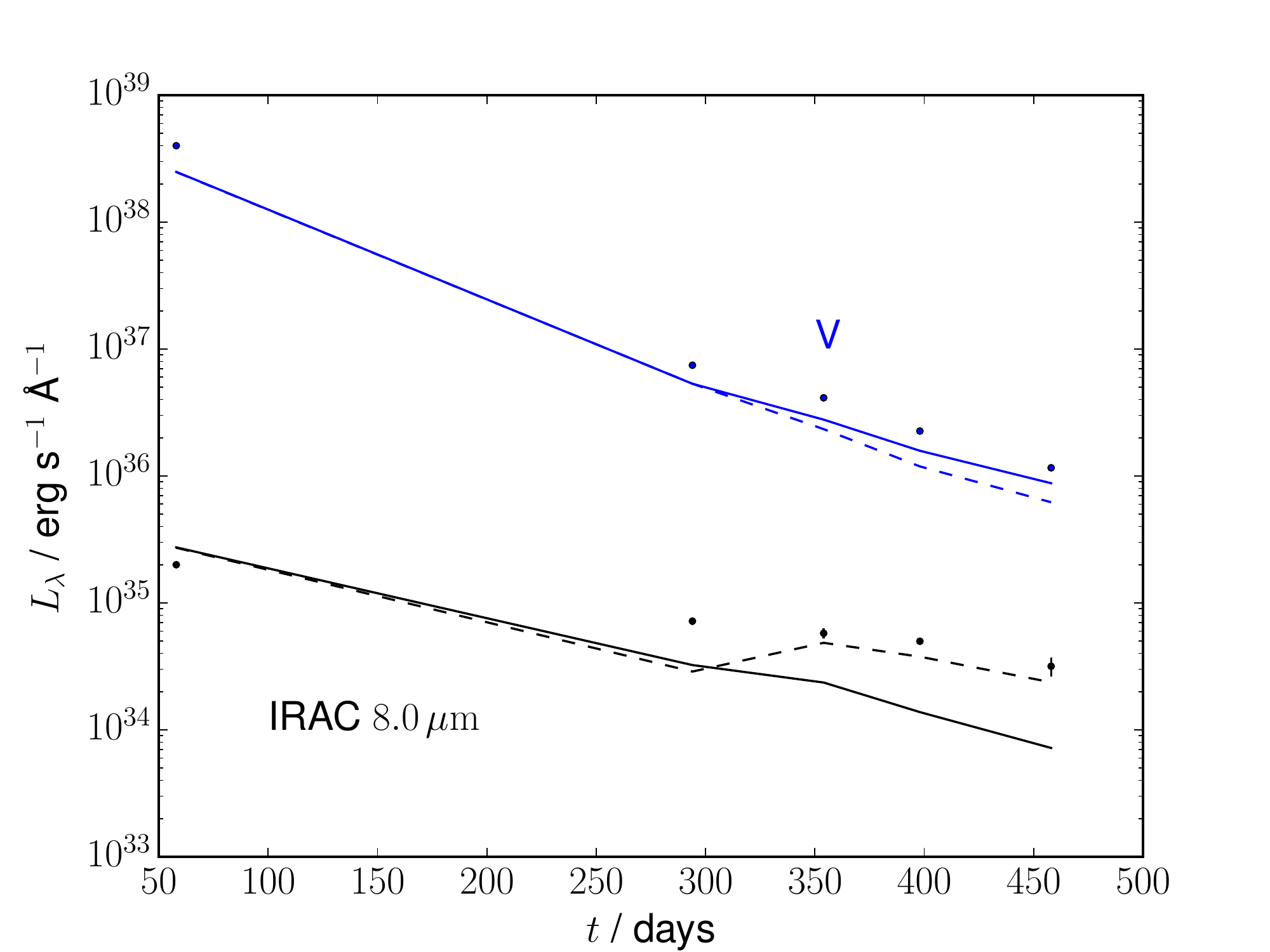}
  \caption{V band (blue) and IRAC $8.0 \um$ (black) fluxes for SN2004et (circles), and { model fluxes} for $0.01 \msun$ of dust with covering fractions $0.1$ (solid lines) and $0.5$ (dashed lines).}
  \label{fig:2004etmodel}
\end{figure}

{ We use as a test case SN2004et, which is well sampled in all photometric bands and shows no signs of interaction, at least at early times ($<500$ days).} Figure \ref{fig:2004etmodel} shows the time evolution of the V band and IRAC $8.0 \um$ fluxes from SN2004et, and from clumped {\sc mocassin} models of $0.01 \msun$ of dust heated by the observed optical-UV flux, with the ejecta assumed to be expanding at a maximum velocity of $3000 \kms$. { Model parameters are given in Table \ref{tab:2004et}.} The model IR fluxes are significantly lower than those observed after a few hundred days for a covering fraction of $0.1$. The fluxes are in better agreement for the higher covering fraction, as a larger surface area is available to emit IR radiation, but this also increases the amount of optical flux absorbed by the dust, to the point where rapid formation of a large dust mass should lead to a detectable drop in the light curve ($\sim 1 \, {\rm mag}$ over a few tens of days on top of the natural decline rate). We also note that the presence of silicate features in the mid-IR spectrum of SN2004et strongly suggests that the dust is not optically thick at these wavelengths for this object.

\begin{table}
  \centering
  \caption{{ Input parameters for {\sc mocassin} models of SN2004et.}}
  \begin{tabular}{cccc}
    \hline
    $t$/day & $r$/pc & $L$/$10^{40}$ erg s$^-1$ & $T$/K \\
    \hline
    58 & $4.9 \times 10^{-4}$ & $10^3$ & $4600$ \\
    298 & $2.5 \times 10^{-3}$ & $30$ & $4000$ \\
    354 & $3.0 \times 10^{-3}$ & $16$ & $4000$ \\
    398 & $3.2 \times 10^{-3}$ & $7.8$ & $4300$ \\
    458 & $3.9 \times 10^{-3}$ & $3.3$ & $5200$ \\
    \hline
  \end{tabular}
  \label{tab:2004et}
\end{table}

\begin{figure*}
  \centering
  \subfigure{\includegraphics[width=\columnwidth]{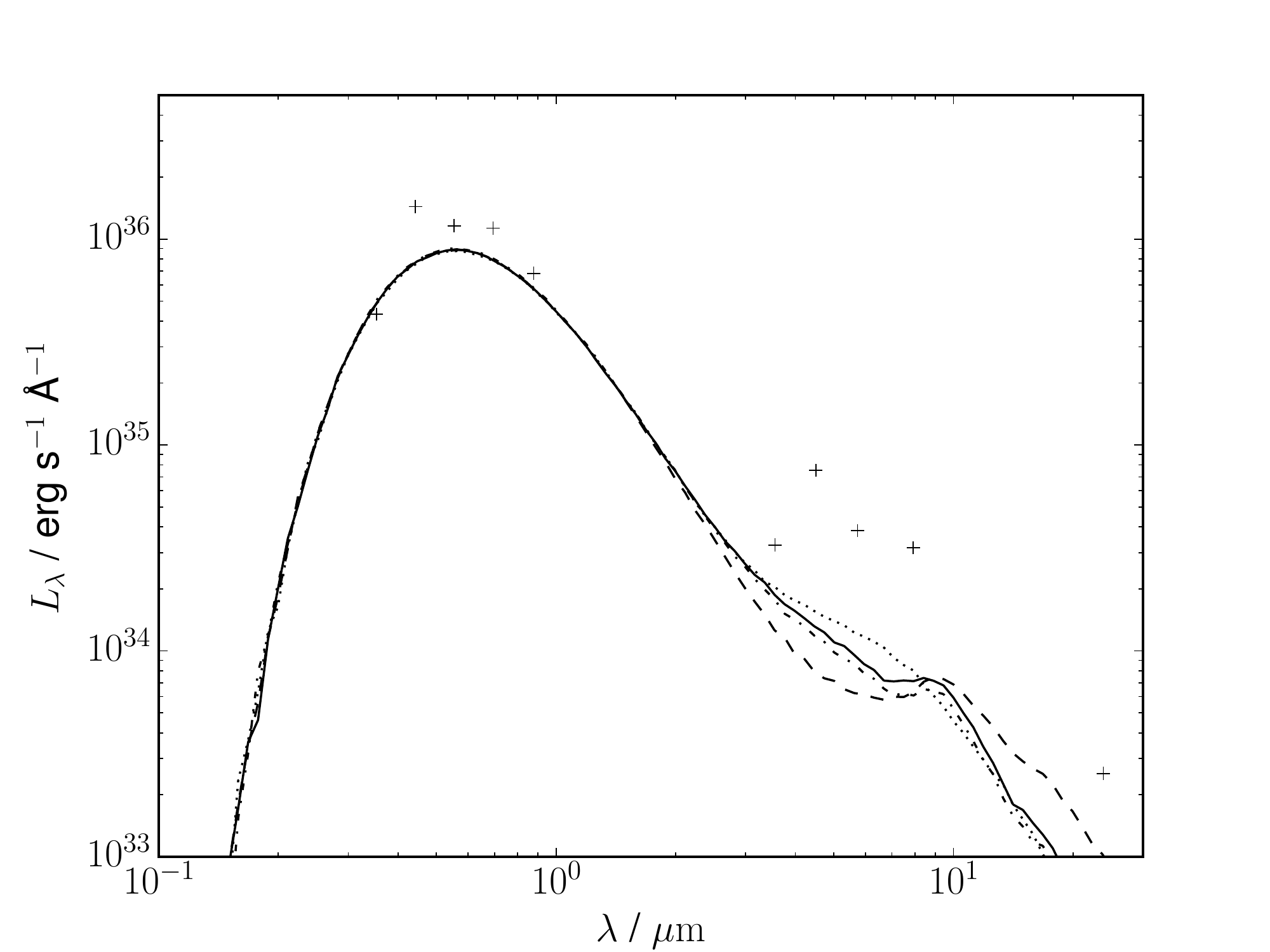}}\quad
  \subfigure{\includegraphics[width=\columnwidth]{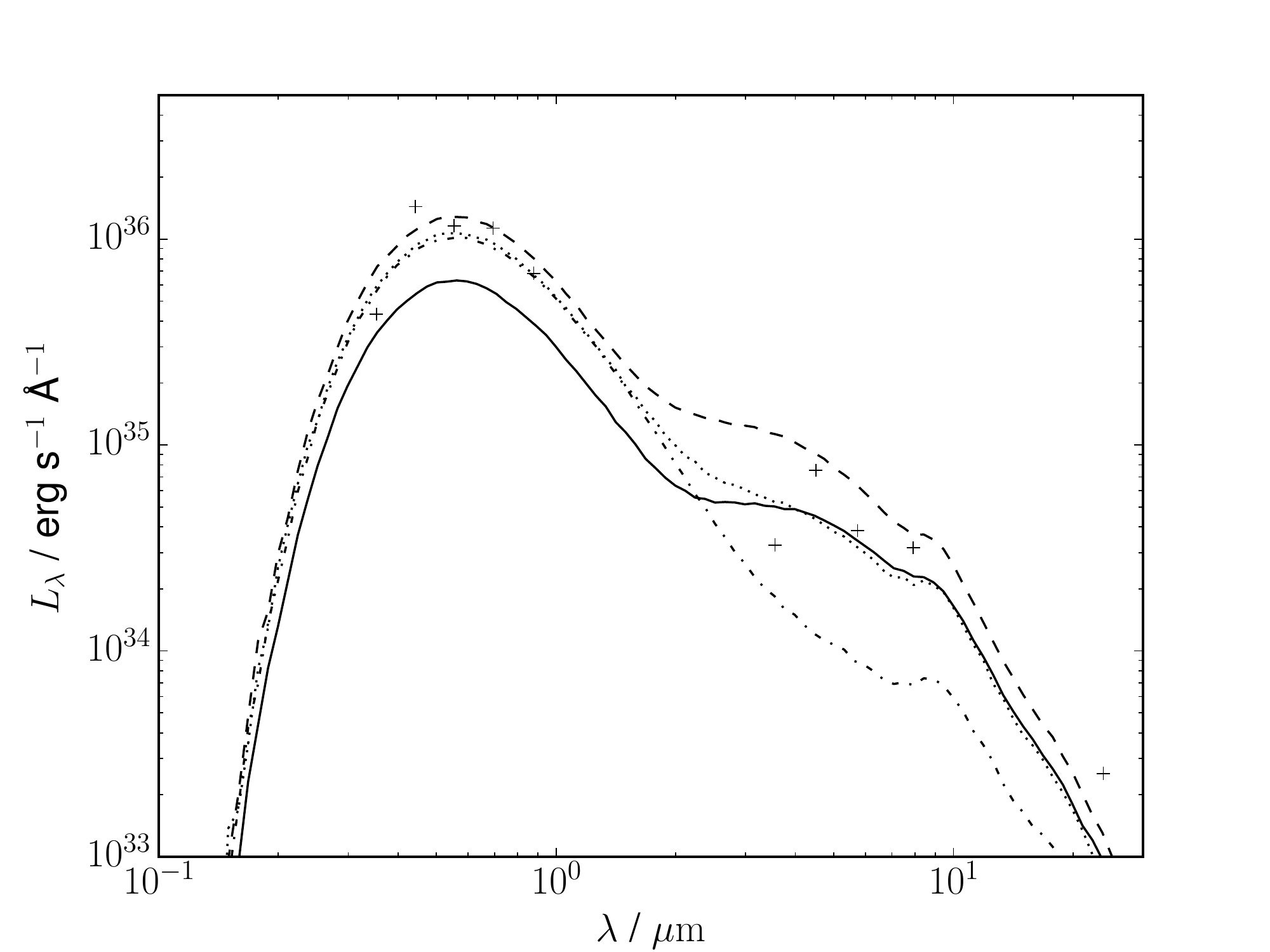}}
  \caption{{ Left panel: observed fluxes of SN2004et at day 458 (black crosses), the fiducial model SED for $f_{\rm cov}=0.1$ (solid line), and those for an increased dust mass of $0.1 \msun$ (dashed line), amorphous carbon dust rather than silicates (dotted line), and a reduced maximum radius of $3.9 \times 10^{-3} \pc$ (dot-dashed line). Right panel: the same observed fluxes (black crosses), with models for the fiducial $f_{\rm cov}=0.5$ model (solid line), and models where the luminosity has been increased by a factor of $f_{\rm cov}^{-1}$ to match the optical photometry for $f_{\rm cov}=0.5$ (dashed line), $0.3$ (dotted line) and $0.1$ (dot-dashed line).}}
  \label{fig:2004ettest}
\end{figure*}

{ Focusing on day 458, we investigate whether varying the model assumptions can reconcile optically thick dust with the observed SED. Figure \ref{fig:2004ettest} shows model SEDs compared to the observed fluxes for our fiducial $f_{\rm cov}=0.1$ model, and ones where we have either increased the dust mass to $0.1 \msun$, reduced the maximum radius to $1.9 \times 10^{-3} \pc$, or replaced the silicate dust with carbon grains using the \citet{zubko1996} BE optical properties. None of these models successfully reproduces the observed IR fluxes, with variations of at most a factor of $\sim 2$. IR fluxes comparable to those observed can only be obtained with higher covering fractions, which additionally require an increased intrinsic luminosity (by a factor of $\sim f_{\rm cov}^{-1}$) to fit the optical photometry, shown in Figure \ref{fig:2004ettest} for $f_{\rm cov}=0.5$, $0.3$ and $0.1$. The $f_{\rm cov}=0.1$ model still fails to reproduce the IR flux, whereas models with larger $f_{\rm cov}$, while consistent with the SED at day 458, would again require a sudden decrease in optical flux as the dust forms which is not seen in the $UBVRI$ light curves.}

\begin{figure*}
  \centering
  \subfigure{\includegraphics[width=\columnwidth]{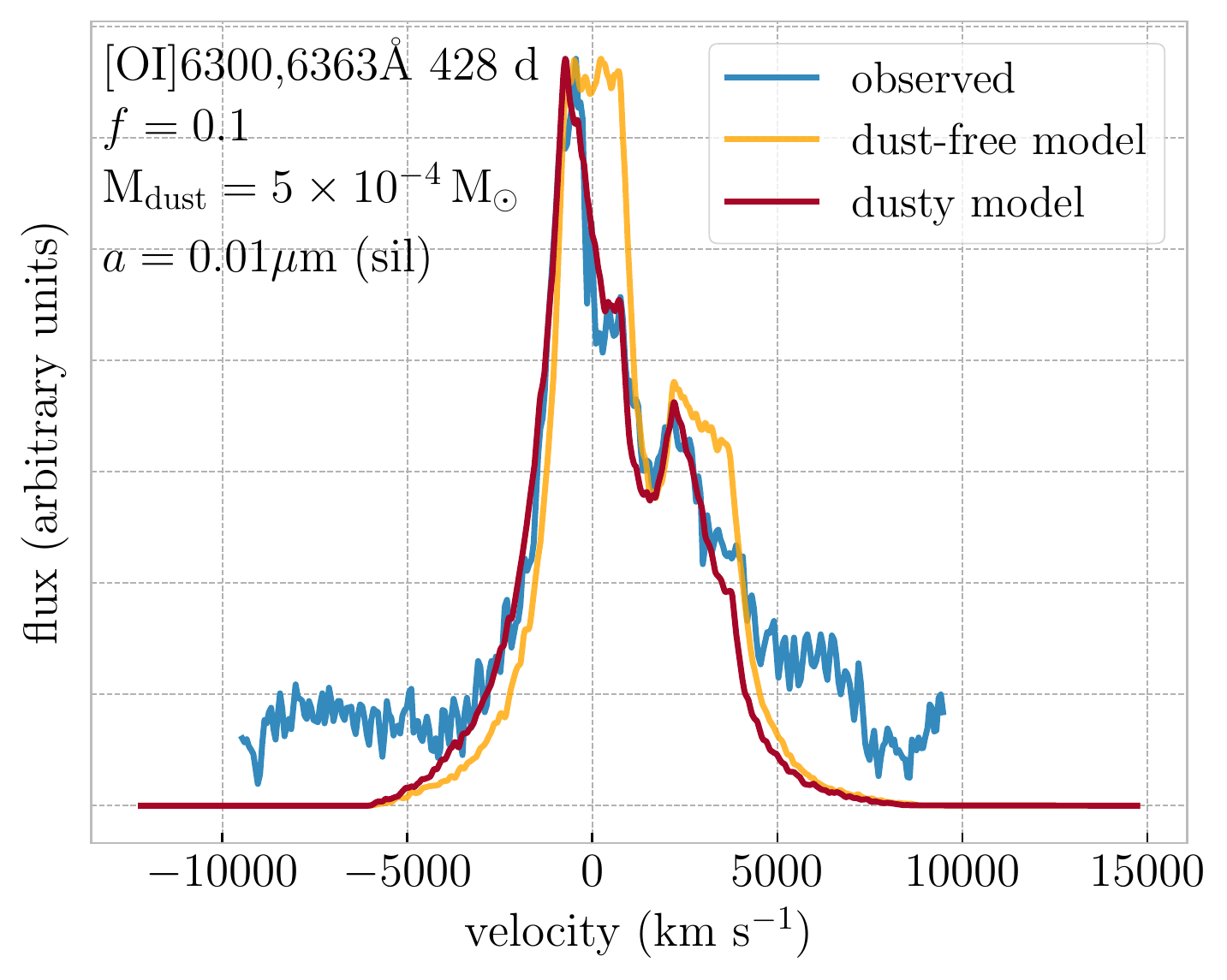}}\quad
  \subfigure{\includegraphics[width=\columnwidth]{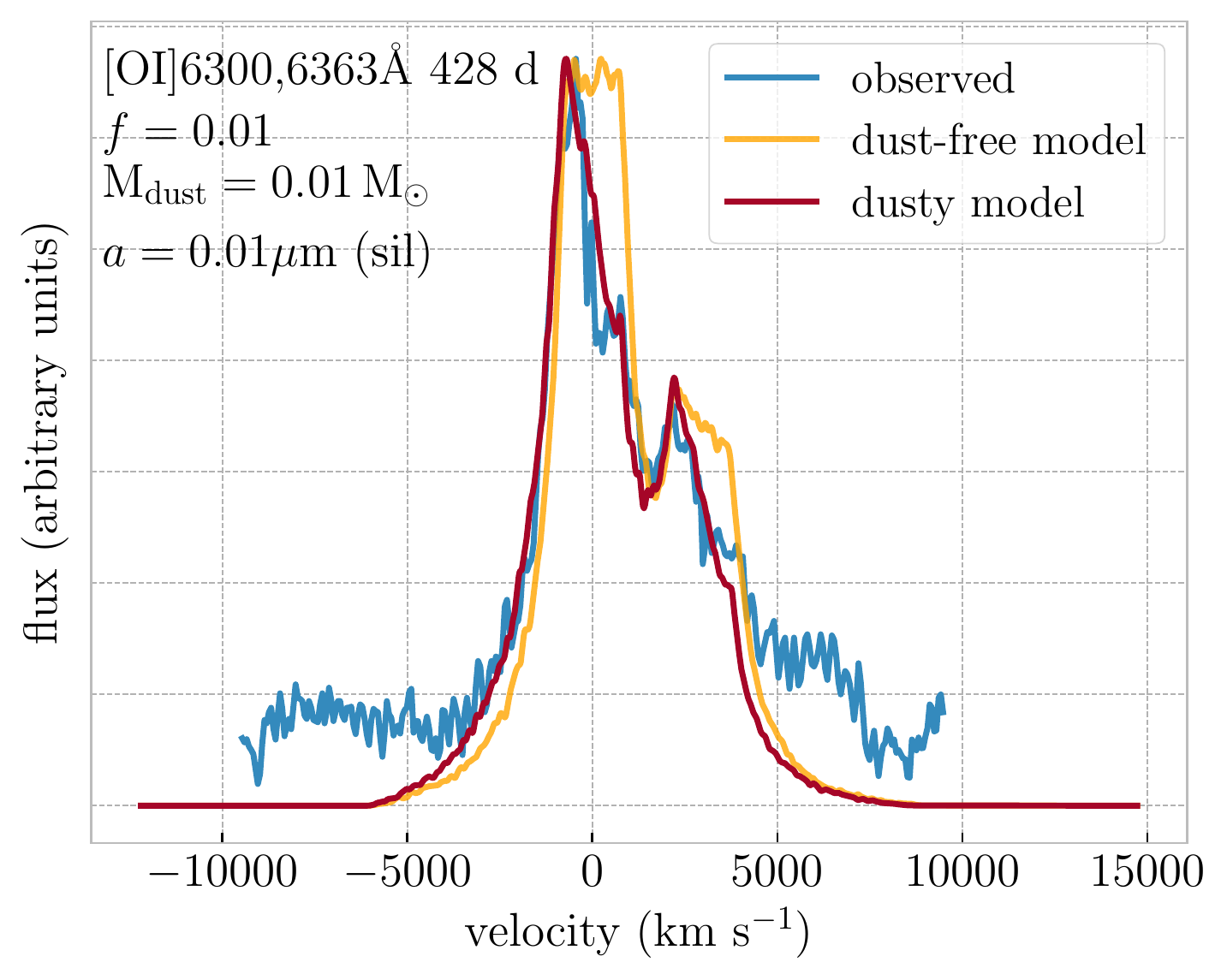}}
  \caption{Observed and model line profiles for the SN2004et [O I] doublet at day 428 \citep{fabbri2011} with and without dust, for $10^{-4} \msun$ of silicate dust with a filling factor of $0.1$ (left panel) and for $0.01 \msun$ of dust with a filling factor of $0.01$ (right panel).}
  \label{fig:lineprofile}
\end{figure*}

In addition to affecting the optical flux, ejecta dust can alter the emission line profiles due to the greater optical depth along the line of sight to receding (red-shifted) ejecta. This can potentially provide an independent constraint on the required degree of clumping. We model the [O I] $6300,6363$~\r{A} doublet from the day 428 spectrum of SN2004et from \citet{fabbri2011} with {\sc damocles} \citep{bevan2016}, using the same assumptions as for the radiative transfer models above. Two representative models are shown in Figure \ref{fig:lineprofile}. A good fit to the observed line profile is possible with $10^{-4} \msun$ of silicate dust for a volume filling factor of $0.1$ (note that this is different to the covering fraction used above) or with a smooth distribution of dust, for a grain size of $0.01 \um$. For $0.01 \msun$ of dust, the line profile can only be reproduced with a filling factor of $0.01$, regardless of the grain size, while larger filling factors are strongly disfavoured. This is because larger filling factors expose the radiation to a larger surface area of dust and therefore require a lower mass, whereas smaller values do not attenuate sufficient radiation. This roughly corresponds to a covering fraction of $0.01^{2/3} \sim 0.05$, smaller than the $0.1$ covering fraction which fails to match the observed IR flux at this epoch. These results - both the line profile fits and the IR emission - also hold if carbon grains, rather than silicates, are used, since for optically thick clumps the grain properties do not greatly affect the radiative transfer. The presence of large dust masses in optically thick clumps requires contradictory clumping properties in order to explain both the optical and IR observations simultaneously.

\begin{figure}
  \centering
  \includegraphics[width=\columnwidth]{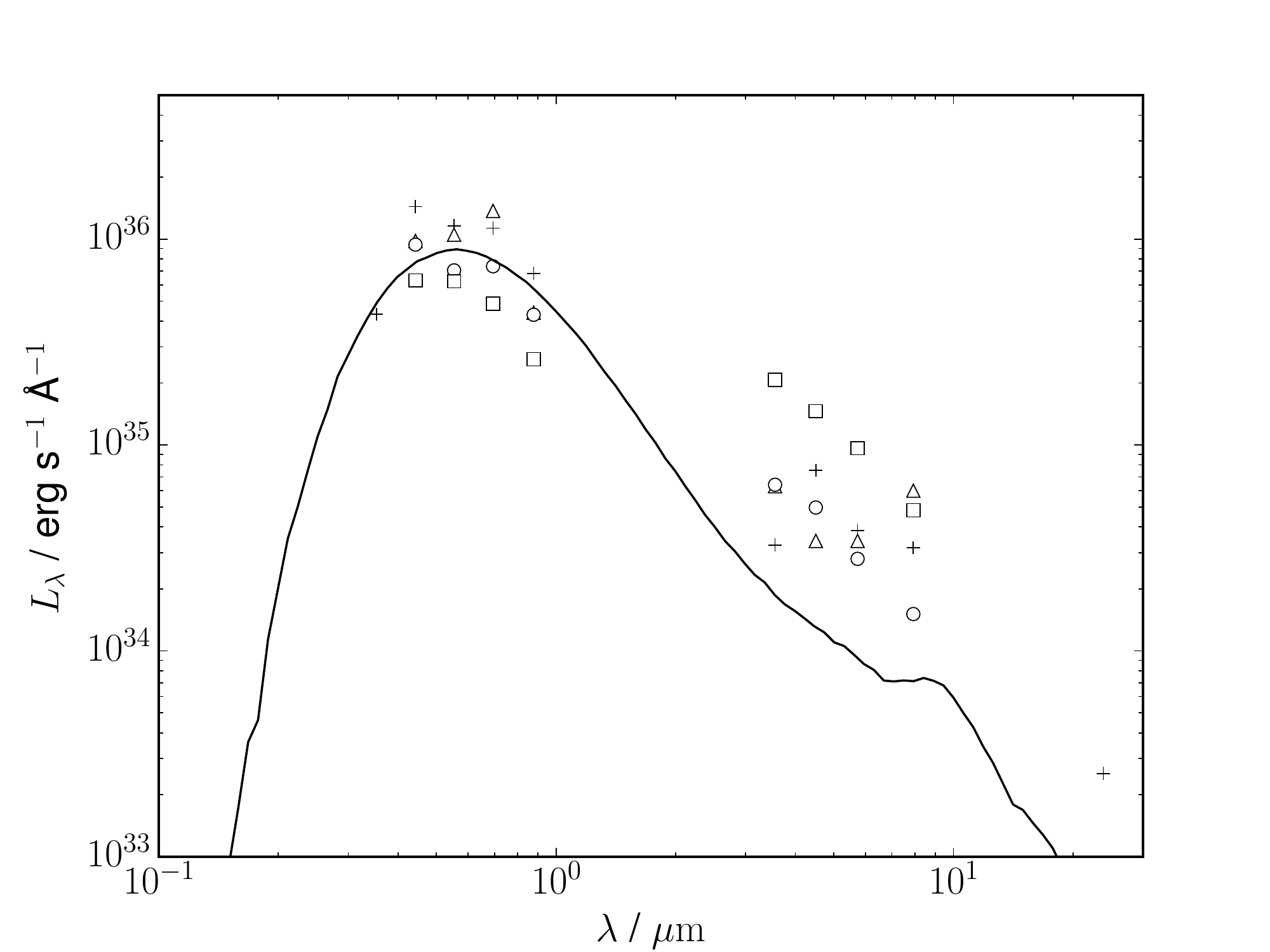}
  \caption{{ Model SED for SN2004et at day 458 with $f = 0.1$ (solid line), and observed fluxes for SN2004A (day 436, triangles), SN2004dj (day 445, circles) and SN2007it (day 560, squares), along with SN2004et at day 458 (crosses).}}
  \label{fig:moresne}
\end{figure}

{ SN2004et is not an exceptional object - Figure \ref{fig:moresne} shows the $f_{\rm cov}=0.1$ model compared to the SEDs of three other SNe (SN2004A, SN2004dj and SN2007it) at similar times. While the optical fluxes are comparable, the model fails to reproduce the IR emission in all cases, despite significant object-to-object variation, which would require correspondingly large changes in the intrinsic luminosity, the covering fraction, or both. We investigate this with a simplified model for the reprocessing of SN radiation by optically thick dust. In the optically thick case, where the absorption cross-section of clumps is equal to the geometric value, the observed optical flux is related to the intrinsic SN luminosity by $L_{\rm opt} = (1-f_{\rm cov})L_0$, while the reprocessed IR luminosity is $L_{\rm IR} = f_{\rm cov}L_0$. $L_{\rm opt}$ and $L_{\rm IR}$ can both be determined from the observed fluxes for each SN in our sample, at least approximately, by integrating the SED over the relevant wavelength ranges, from the U band to $3.6 \um$ for the `optical' and from $3.6 \um$ to the longest wavelength available for the IR. We integrate over the observed fluxes, rather than e.g. blackbody fits to the fluxes, in order to avoid introducing model dependencies. While the true luminosities will depend on the underlying SEDs for the SN and dust, given the previously discussed uncertainties in determining these from photometric data, we consider the integrated fluxes to be acceptable estimates. We can then calculate $L_0$ and $f_{\rm cov}$ by trivially rearranging the two previous equations.}

\begin{figure*}
  \centering
  \subfigure{\includegraphics[width=\columnwidth]{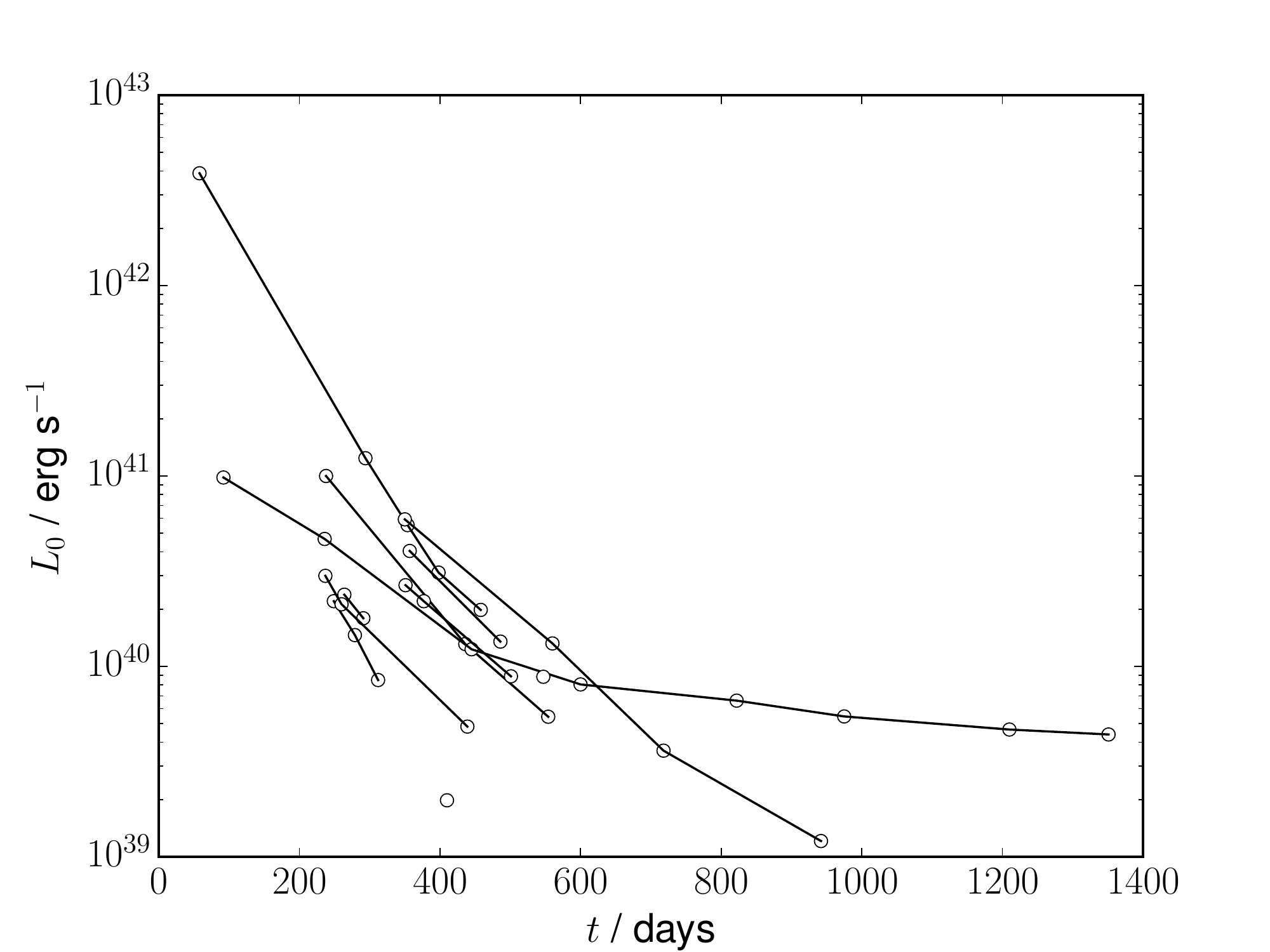}}\quad
  \subfigure{\includegraphics[width=\columnwidth]{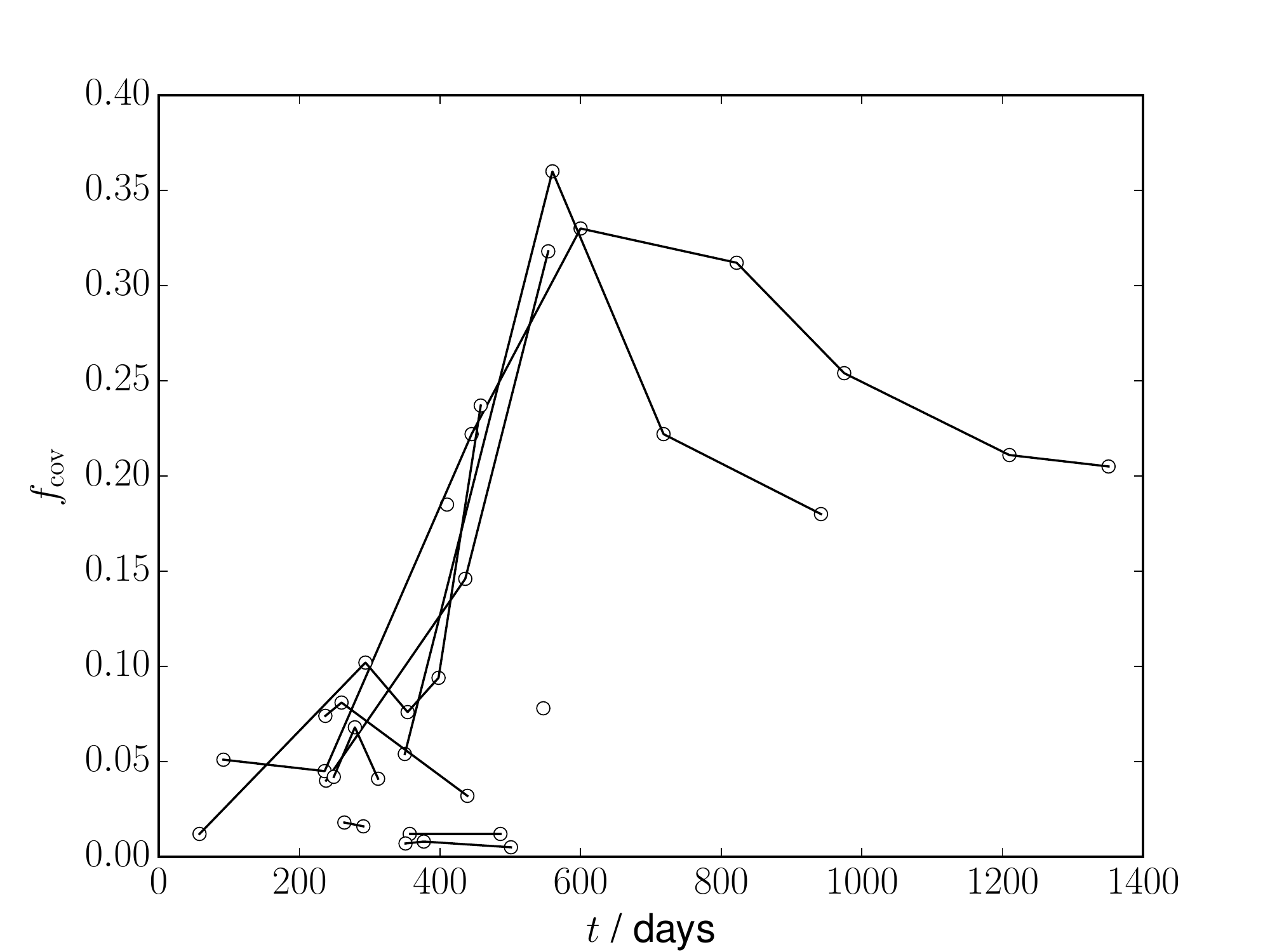}}
  \caption{{ Intrinsic luminosity $L_0$ (left) and covering fraction $f_{\rm cov}$ (right) derived from observed fluxes for each supernova as a function of time. Individual epochs are marked as circles, and those from the same supernova are linked by solid lines.}}
  \label{fig:blahplot}
\end{figure*}

{ Figure \ref{fig:blahplot} shows the derived values of $L_0$ and $f_{\rm cov}$ for each epoch. The values of $L_0$ appear reasonable, declining over time for each SN as would be expected, although with significant object-to-object diversity. The covering fraction also varies between SNe, from $\sim 0.01-0.35$, and not only changes with time for individual objects but does so non-monotonically. This is inconsistent with a constant number of clumps expanding at the same rate as the ejecta, as assumed by \citet{dwek2019}. An increasing $f_{\rm cov}$ could be explained by the ongoing formation of new clumps, but the subsequent decline seen in several objects would then require that the combined surface area of the clumps falls relative to the ejecta. This would be the case for ballistically expanding clumps, but $f_{\rm cov}$ should then fall as $\sim t^{-2}$, whereas the actual declines are much slower than this. A consistent physical model for the evolution of $f_{\rm cov}$, assuming one can be found, would have to be far more complicated than the ones proposed by \citet{dwek2015} and \citet{dwek2019}. A detailed radiative transfer study of dust clumping in SN1987A (Bevan \& Wesson, subm.) does not find any such model with large dust masses that reproduces the combined optical-IR evolution of that object. While optically thick clumps are effectively impossible to rule out, as $f_{\rm cov}$ can be made arbitrarily small, as a natural explanation for the observed properties of CCSNe they clearly fail.}

\section{Conclusions}

{ Determining the dust mass in SNe from the IR emission can be affected by several systematic issues, which may result in simple blackbody fits returning erroneously low values. We have shown, using a method that removes the most significant of these effects, that the resulting masses inferred for a sample of 11 CCSNe are still orders of magnitude below those predicted by dust formation models at early times. These large dust masses, if present, must be optically thick in the IR to be consistent with observations. Considering both broad-band and line emission at optical wavelengths, the covering fraction of this optically thick dust has to be $\lesssim 0.1$ to avoid observational effects which are not seen in SNe. Conversely, the observed IR fluxes require typical covering fractions $\gtrsim 0.2$. Moreover, the diversity in IR emission between SNe with similar optical properties implies that each object requires a different covering fraction, even assuming a value can be found to simultaneously reproduce the optical and IR properties. A simple analysis suggests that this covering fraction must vary, non-monotonically, with time for at least some SNe. While more complex models may be able to achieve this, large masses of optically thick dust appear to require significant finetuning to remain consistent with observations of CCSNe.}

{ The difficulty in finding a consistent model for optically thick dust contrasts with the relative ease of fitting the data with low dust masses. Even previous studies using full radiative transfer models (e.g. \citealt{fabbri2011,wesson2015}) prefer early-time dust masses $\lesssim 10^{-3} \msun$, due to the difficulty of simultaneously fitting optical and IR data without invoking arbitrary changes in the underlying SN luminosity. Using two entirely independent methods, \citet{wesson2015} and \citet{bevan2016} obtained consistent results for the dust mass evolution in SN1987A, with a small but non-zero dust mass forming within $\sim 1000$ days, followed by an increase over the next several decades to the large ($>0.1 \msun$) values detected in the present day \citep{matsuura2015,cigan2019}. This picture of dust formation in CCSNe is supported by all other observational evidence \citep{gall2014,szalai2019,bevan2019}. Current theoretical models essentially require that large masses of dust form, but remain undetectable by any method for years to decades afterwards, in every observed CCSNe to date. We consider it at least plausible that instead, a mechanism for late-time dust formation in CCSNe exists and has so far been neglected.}

\section*{Acknowledgements}

{ We are grateful to the referee for suggestions which significantly improved the content and focus of this paper.} FDP is funded by the Science and Technology Facilities Council. AB and MJB acknowledge support from the European Research Council (ERC) grant SNDUST ERC-2015-AdG-694520. IDL gratefully acknowledges the support of the Research Foundation -- Flanders (FWO).

\section*{Data Availability}

{ The data underlying this article are publically available at either the OSC ({\url www.sne.space}) or the references listed in Table \ref{tab:photometry}. The derived data used in the analysis are tabulated in the Appendix. The codes used are all publically available, at: {\url www.github.com/fpriestley/dinamo} (\dinamo{}); {\url www.mocassin.nebulousresearch.org} ({\sc mocassin}); {\url www.github.com/damocles-code/damocles} ({\sc damocles}).}




\bibliographystyle{mnras}
\bibliography{lightcurve}


\appendix

\section{Table of photometric data}

Table \ref{tab:sed} lists the photometric data used or obtained for each epoch, from the references given in Table \ref{tab:photometry}, and Table \ref{tab:sederr} lists the uncertainties.

\begin{table*}
  \centering
  \caption{SEDs for each SN and epoch. Luminosities are in erg s$^{-1}$ \r{A}$^{-1}$. References can be found in Table \ref{tab:photometry}.}
  \begin{tabular}{cccccccccccc}
    \hline
    SN & Day & U & B & V & R & I & $3.6 \um$ & $4.5 \um$ & $5.8 \um$ & $8.0 \um$ & $24 \um$ \\
    \hline
2003gd & 410 & 0.0e+00 & 1.6e+35 & 1.4e+35 & 1.3e+35 & 7.5e+34 & 5.1e+33 & 1.1e+34 & 7.9e+33 & 7.7e+33 & 0.0e+00 \\ 
2004A & 238 & 0.0e+00 & 4.4e+36 & 6.0e+36 & 7.4e+36 & 5.0e+36 & 1.5e+35 & 1.3e+35 & 6.2e+34 & 7.9e+34 & 0.0e+00 \\ 
2004A & 436 & 0.0e+00 & 9.8e+35 & 1.1e+36 & 1.4e+36 & 4.4e+35 & 6.3e+34 & 3.4e+34 & 3.4e+34 & 6.0e+34 & 0.0e+00 \\ 
2004A & 554 & 0.0e+00 & 4.3e+35 & 3.8e+35 & 5.4e+35 & 1.0e+35 & 4.8e+34 & 2.4e+34 & 3.8e+34 & 5.4e+34 & 0.0e+00 \\ 
2004dj & 1210 & 0.0e+00 & 5.0e+35 & 3.6e+35 & 2.3e+35 & 1.7e+35 & 1.2e+34 & 9.3e+33 & 9.5e+33 & 6.5e+33 & 9.2e+32 \\ 
2004dj & 1351 & 0.0e+00 & 5.2e+35 & 3.6e+35 & 2.3e+35 & 1.6e+35 & 1.1e+34 & 7.9e+33 & 8.0e+33 & 6.0e+33 & 9.5e+32 \\ 
2004dj & 236 & 0.0e+00 & 2.2e+36 & 2.4e+36 & 3.4e+36 & 2.4e+36 & 4.3e+34 & 7.4e+34 & 1.9e+34 & 8.1e+33 & 5.7e+32 \\ 
2004dj & 445 & 0.0e+00 & 9.4e+35 & 7.1e+35 & 7.4e+35 & 4.3e+35 & 6.4e+34 & 5.0e+34 & 2.8e+34 & 1.5e+34 & 6.1e+32 \\ 
2004dj & 600 & 0.0e+00 & 6.1e+35 & 4.1e+35 & 3.7e+35 & 2.3e+35 & 4.6e+34 & 3.7e+34 & 2.9e+34 & 1.6e+34 & 7.6e+32 \\ 
2004dj & 822 & 0.0e+00 & 5.0e+35 & 3.5e+35 & 2.4e+35 & 2.1e+35 & 2.9e+34 & 2.4e+34 & 2.2e+34 & 1.4e+34 & 8.8e+32 \\ 
2004dj & 92 & 1.1e+36 & 4.4e+36 & 7.8e+36 & 6.3e+36 & 4.6e+36 & 1.5e+35 & 1.6e+35 & 5.8e+34 & 1.6e+34 & 5.3e+32 \\ 
2004dj & 975 & 0.0e+00 & 4.6e+35 & 3.4e+35 & 2.4e+35 & 1.9e+35 & 1.7e+34 & 1.4e+34 & 1.4e+34 & 9.5e+33 & 6.9e+32 \\ 
2004et & 294 & 5.5e+35 & 1.8e+36 & 2.8e+36 & 3.3e+36 & 2.1e+36 & 5.1e+34 & 1.3e+35 & 3.0e+34 & 2.7e+34 & 1.2e+33 \\ 
2004et & 354 & 3.5e+35 & 1.2e+36 & 1.5e+36 & 1.7e+36 & 8.7e+35 & 3.0e+34 & 7.4e+34 & 2.2e+34 & 2.1e+34 & 0.0e+00 \\ 
2004et & 398 & 2.5e+35 & 8.3e+35 & 8.4e+35 & 9.2e+35 & 4.7e+35 & 2.2e+34 & 4.5e+34 & 1.7e+34 & 1.9e+34 & 0.0e+00 \\ 
2004et & 458 & 1.6e+35 & 5.4e+35 & 4.3e+35 & 4.2e+35 & 2.5e+35 & 1.2e+34 & 2.8e+34 & 1.4e+34 & 1.2e+34 & 9.5e+32 \\ 
2004et & 58 & 1.5e+37 & 9.4e+37 & 1.5e+38 & 1.1e+38 & 6.6e+37 & 1.2e+36 & 5.6e+35 & 2.4e+35 & 7.5e+34 & 0.0e+00 \\ 
2007it & 350 & 0.0e+00 & 3.8e+36 & 4.8e+36 & 4.9e+36 & 2.7e+36 & 9.6e+34 & 1.5e+35 & 4.6e+34 & 2.8e+34 & 0.0e+00 \\ 
2007it & 560 & 0.0e+00 & 6.3e+35 & 6.2e+35 & 4.9e+35 & 2.6e+35 & 2.1e+35 & 1.5e+35 & 9.7e+34 & 4.8e+34 & 0.0e+00 \\ 
2007it & 718 & 0.0e+00 & 2.7e+35 & 2.1e+35 & 1.9e+35 & 6.8e+34 & 8.3e+34 & 8.6e+34 & 0.0e+00 & 0.0e+00 & 0.0e+00 \\ 
2007it & 942 & 0.0e+00 & 2.0e+35 & 1.1e+35 & 7.0e+34 & 3.1e+34 & 1.4e+34 & 3.2e+34 & 0.0e+00 & 0.0e+00 & 0.0e+00 \\ 
2009E & 547 & 0.0e+00 & 9.6e+35 & 6.6e+35 & 7.6e+35 & 3.1e+35 & 4.1e+34 & 9.4e+34 & 0.0e+00 & 0.0e+00 & 0.0e+00 \\ 
2011dh & 249 & 0.0e+00 & 2.2e+36 & 2.2e+36 & 1.5e+36 & 9.4e+35 & 9.6e+34 & 9.8e+34 & 0.0e+00 & 0.0e+00 & 0.0e+00 \\ 
2011dh & 279 & 0.0e+00 & 1.5e+36 & 1.5e+36 & 9.1e+35 & 5.8e+35 & 8.2e+34 & 1.3e+35 & 0.0e+00 & 0.0e+00 & 0.0e+00 \\ 
2011dh & 312 & 0.0e+00 & 1.0e+36 & 1.0e+36 & 6.2e+35 & 3.4e+35 & 3.2e+34 & 4.2e+34 & 0.0e+00 & 0.0e+00 & 0.0e+00 \\ 
2012aw & 357 & 2.6e+35 & 1.3e+36 & 2.2e+36 & 2.6e+36 & 2.2e+36 & 3.0e+34 & 7.4e+34 & 0.0e+00 & 0.0e+00 & 0.0e+00 \\ 
2012aw & 486 & 1.9e+35 & 4.4e+35 & 7.4e+35 & 6.4e+35 & 7.5e+35 & 1.1e+34 & 2.5e+34 & 0.0e+00 & 0.0e+00 & 0.0e+00 \\ 
2013am & 351 & 0.0e+00 & 0.0e+00 & 5.8e+35 & 1.2e+36 & 1.7e+36 & 2.4e+34 & 1.4e+34 & 0.0e+00 & 0.0e+00 & 0.0e+00 \\ 
2013am & 377 & 0.0e+00 & 0.0e+00 & 5.4e+35 & 1.1e+36 & 1.4e+36 & 1.9e+34 & 2.0e+34 & 0.0e+00 & 0.0e+00 & 0.0e+00 \\ 
2013am & 501 & 0.0e+00 & 0.0e+00 & 3.8e+35 & 8.2e+35 & 5.0e+35 & 2.6e+33 & 7.4e+33 & 0.0e+00 & 0.0e+00 & 0.0e+00 \\ 
2013df & 264 & 0.0e+00 & 1.4e+36 & 1.2e+36 & 1.0e+36 & 1.3e+36 & 3.8e+34 & 5.1e+34 & 0.0e+00 & 0.0e+00 & 0.0e+00 \\ 
2013df & 291 & 0.0e+00 & 6.5e+35 & 1.0e+36 & 8.7e+35 & 1.0e+36 & 2.2e+34 & 3.8e+34 & 0.0e+00 & 0.0e+00 & 0.0e+00 \\ 
2013ej & 237 & 0.0e+00 & 1.8e+36 & 2.7e+36 & 3.0e+36 & 1.1e+36 & 2.2e+35 & 2.5e+35 & 0.0e+00 & 0.0e+00 & 0.0e+00 \\ 
2013ej & 260 & 0.0e+00 & 1.5e+36 & 2.1e+36 & 2.2e+36 & 7.1e+35 & 1.7e+35 & 2.0e+35 & 0.0e+00 & 0.0e+00 & 0.0e+00 \\ 
2013ej & 439 & 0.0e+00 & 4.1e+35 & 3.1e+35 & 2.6e+35 & 0.0e+00 & 1.1e+34 & 2.2e+34 & 0.0e+00 & 0.0e+00 & 0.0e+00 \\
\hline
  \end{tabular}
  \label{tab:sed}
\end{table*}

\begin{table*}
  \centering
  \caption{Uncertainties for each SN and epoch in erg s$^{-1}$ \r{A}$^{-1}$. References can be found in Table \ref{tab:photometry}.}
  \begin{tabular}{cccccccccccc}
    \hline
    SN & Day & U & B & V & R & I & $3.6 \um$ & $4.5 \um$ & $5.8 \um$ & $8.0 \um$ & $24 \um$ \\
    \hline
2003gd & 410 & 0.0e+00 & 8.7e+33 & 1.4e+34 & 1.1e+34 & 8.2e+33 & 2.3e+33 & 2.3e+33 & 6.0e+32 & 7.3e+32 & 0.0e+00 \\ 
2004A & 238 & 0.0e+00 & 3.3e+35 & 3.7e+35 & 6.6e+35 & 1.6e+35 & 5.6e+34 & 4.8e+34 & 1.1e+34 & 9.3e+33 & 0.0e+00 \\ 
2004A & 436 & 0.0e+00 & 4.1e+35 & 7.8e+34 & 6.9e+35 & 4.8e+34 & 2.6e+34 & 1.4e+34 & 9.3e+33 & 8.8e+33 & 0.0e+00 \\ 
2004A & 554 & 0.0e+00 & 1.8e+35 & 2.8e+34 & 9.9e+34 & 1.6e+34 & 2.2e+34 & 1.1e+34 & 9.3e+33 & 9.3e+33 & 0.0e+00 \\ 
2004dj & 1210 & 0.0e+00 & 1.9e+34 & 1.6e+34 & 6.4e+33 & 1.7e+34 & 3.0e+33 & 2.3e+33 & 6.6e+32 & 6.5e+32 & 4.6e+31 \\ 
2004dj & 1351 & 0.0e+00 & 2.2e+34 & 1.6e+34 & 6.4e+33 & 1.7e+34 & 2.8e+33 & 2.0e+33 & 8.8e+32 & 6.7e+32 & 6.9e+31 \\ 
2004dj & 236 & 0.0e+00 & 8.2e+34 & 9.0e+34 & 9.4e+34 & 6.6e+34 & 1.1e+34 & 1.9e+34 & 3.4e+33 & 8.9e+32 & 1.3e+32 \\ 
2004dj & 445 & 0.0e+00 & 3.8e+34 & 2.8e+34 & 2.7e+34 & 2.5e+34 & 1.6e+34 & 1.2e+34 & 5.9e+33 & 2.0e+33 & 1.5e+32 \\ 
2004dj & 600 & 0.0e+00 & 4.7e+34 & 4.1e+34 & 2.3e+34 & 1.8e+34 & 1.1e+34 & 9.2e+33 & 4.6e+33 & 2.0e+33 & 1.5e+32 \\ 
2004dj & 822 & 0.0e+00 & 2.0e+34 & 9.9e+33 & 6.9e+33 & 1.6e+34 & 7.1e+33 & 6.0e+33 & 2.8e+33 & 1.6e+33 & 1.1e+32 \\ 
2004dj & 92 & 9.7e+34 & 1.2e+35 & 1.4e+35 & 2.4e+35 & 9.8e+34 & 3.7e+34 & 4.1e+34 & 1.5e+34 & 1.8e+33 & 6.9e+31 \\ 
2004dj & 975 & 0.0e+00 & 1.3e+34 & 9.3e+33 & 6.5e+33 & 1.7e+34 & 4.1e+33 & 3.5e+33 & 2.0e+33 & 1.1e+33 & 7.6e+31 \\ 
2004et & 294 & 7.2e+34 & 7.6e+34 & 1.2e+35 & 1.1e+35 & 9.8e+34 & 3.8e+33 & 1.7e+34 & 1.4e+34 & 5.1e+33 & 2.8e+32 \\ 
2004et & 354 & 2.8e+34 & 1.5e+35 & 1.9e+35 & 1.4e+35 & 9.3e+34 & 3.7e+33 & 9.1e+33 & 1.8e+34 & 5.2e+33 & 0.0e+00 \\ 
2004et & 398 & 2.8e+34 & 1.2e+35 & 7.9e+34 & 6.5e+34 & 1.4e+35 & 1.6e+33 & 5.6e+33 & 5.0e+33 & 2.6e+33 & 0.0e+00 \\ 
2004et & 458 & 2.9e+34 & 2.6e+34 & 1.9e+34 & 1.4e+34 & 1.6e+34 & 1.8e+33 & 2.1e+33 & 1.9e+34 & 5.4e+33 & 3.3e+32 \\ 
2004et & 58 & 4.4e+36 & 4.4e+36 & 3.7e+36 & 2.8e+36 & 1.6e+36 & 9.1e+34 & 6.9e+34 & 1.1e+34 & 3.0e+33 & 0.0e+00 \\ 
2007it & 350 & 0.0e+00 & 2.2e+35 & 1.8e+35 & 1.8e+35 & 1.2e+35 & 1.1e+34 & 8.2e+33 & 1.9e+33 & 1.8e+33 & 0.0e+00 \\ 
2007it & 560 & 0.0e+00 & 8.2e+34 & 2.3e+34 & 1.7e+34 & 1.1e+34 & 1.1e+34 & 8.1e+33 & 2.6e+33 & 1.8e+33 & 0.0e+00 \\ 
2007it & 718 & 0.0e+00 & 3.2e+34 & 9.1e+33 & 5.2e+33 & 3.1e+33 & 9.2e+33 & 7.9e+33 & 0.0e+00 & 0.0e+00 & 0.0e+00 \\ 
2007it & 942 & 0.0e+00 & 3.4e+34 & 2.9e+33 & 1.9e+33 & 1.1e+33 & 5.5e+33 & 5.0e+33 & 0.0e+00 & 0.0e+00 & 0.0e+00 \\ 
2009E & 547 & 0.0e+00 & 5.0e+35 & 2.2e+35 & 2.5e+35 & 5.4e+34 & 1.6e+34 & 9.6e+33 & 0.0e+00 & 0.0e+00 & 0.0e+00 \\ 
2011dh & 249 & 0.0e+00 & 3.8e+35 & 4.5e+35 & 6.6e+34 & 1.1e+35 & 8.9e+33 & 9.0e+33 & 0.0e+00 & 0.0e+00 & 0.0e+00 \\ 
2011dh & 279 & 0.0e+00 & 3.5e+35 & 4.2e+35 & 5.5e+34 & 8.5e+34 & 7.6e+33 & 1.2e+34 & 0.0e+00 & 0.0e+00 & 0.0e+00 \\ 
2011dh & 312 & 0.0e+00 & 3.0e+35 & 3.7e+35 & 2.1e+34 & 6.3e+34 & 2.9e+33 & 3.9e+33 & 0.0e+00 & 0.0e+00 & 0.0e+00 \\ 
2012aw & 357 & 1.8e+34 & 4.3e+34 & 4.0e+34 & 2.4e+34 & 2.9e+34 & 2.8e+33 & 6.8e+33 & 0.0e+00 & 0.0e+00 & 0.0e+00 \\ 
2012aw & 486 & 9.7e+33 & 2.6e+34 & 1.4e+34 & 5.6e+34 & 8.8e+33 & 1.9e+33 & 2.3e+33 & 0.0e+00 & 0.0e+00 & 0.0e+00 \\ 
2013am & 351 & 0.0e+00 & 0.0e+00 & 3.4e+34 & 1.7e+35 & 1.8e+35 & 9.7e+33 & 5.5e+33 & 0.0e+00 & 0.0e+00 & 0.0e+00 \\ 
2013am & 377 & 0.0e+00 & 0.0e+00 & 1.4e+34 & 2.1e+35 & 1.9e+35 & 8.7e+33 & 6.7e+33 & 0.0e+00 & 0.0e+00 & 0.0e+00 \\ 
2013am & 501 & 0.0e+00 & 0.0e+00 & 4.9e+34 & 3.2e+35 & 1.5e+35 & 2.7e+33 & 4.0e+33 & 0.0e+00 & 0.0e+00 & 0.0e+00 \\ 
2013df & 264 & 0.0e+00 & 1.0e+35 & 8.7e+34 & 7.4e+34 & 7.0e+34 & 1.2e+34 & 1.6e+34 & 0.0e+00 & 0.0e+00 & 0.0e+00 \\ 
2013df & 291 & 0.0e+00 & 4.8e+34 & 6.7e+34 & 5.7e+34 & 4.9e+34 & 8.8e+33 & 1.3e+34 & 0.0e+00 & 0.0e+00 & 0.0e+00 \\ 
2013ej & 237 & 0.0e+00 & 1.4e+35 & 2.9e+35 & 1.9e+35 & 2.2e+35 & 2.6e+34 & 1.6e+34 & 0.0e+00 & 0.0e+00 & 0.0e+00 \\ 
2013ej & 260 & 0.0e+00 & 1.2e+35 & 2.2e+35 & 1.5e+35 & 1.9e+35 & 2.0e+34 & 9.1e+33 & 0.0e+00 & 0.0e+00 & 0.0e+00 \\ 
2013ej & 439 & 0.0e+00 & 2.9e+34 & 3.2e+34 & 2.3e+34 & 0.0e+00 & 1.8e+33 & 2.4e+33 & 0.0e+00 & 0.0e+00 & 0.0e+00 \\
\hline
  \end{tabular}
  \label{tab:sederr}
\end{table*}


\bsp	
\label{lastpage}
\end{document}